\documentstyle[11pt,moriond]{article}
\input{psfig}

\def\Journal#1#2#3#4{{#1} {\bf #2}, #3 (#4)}

\def\NPB{{\em Nucl. Phys. B}}
\def\PLB{{\em Phys. Lett.  B}}
\def\PRL{\em Phys. Rev. Lett.}
\def\PRD{{\em Phys. Rev. D}}

\def\APJ{{\em Astrophys. J.}}
\def\APJL{{\em Astrophys. J. Lett.}}

\def\MNRASL{{\em Mon. Not. R. Astron. Soc.}}

\def\be{\begin{equation}}
\def\ee{\end{equation}}
\def\bea{\begin{eqnarray}}
\def\eea{\end{eqnarray}}

\def\VEV#1{\left\langle #1\right\rangle}
\def\sec{\ifmmode \,\, {\rm sec} \else sec \fi}
\def\eV {\ifmmode \,\, {\rm eV} \else eV \fi}
\def\keV{\ifmmode \,\, {\rm keV} \else keV \fi}
\def\MeV{\ifmmode \,\, {\rm MeV} \else MeV \fi}
\def\GeV{\ifmmode \,\, {\rm GeV} \else GeV \fi}
\def\TeV{\ifmmode \,\, {\rm TeV} \else TeV \fi}
\def\fm{\ifmmode \,\, {\rm fm} \else TeV \fi}
\def\pbarn{\ifmmode \,\, {\rm pb} \else pb \fi}
\def\km{\ifmmode {\rm km}\, \else km \fi}
\def\Mpc{\ifmmode {\rm Mpc}\, \else Mpc \fi}
\def\Gyr{\ifmmode {\rm Gyr}\, \else Gyr \fi}

\def\fun#1#2{\lower3.6pt\vbox{\baselineskip0pt\lineskip.9pt
  \ialign{$\mathsurround=0pt#1\hfil##\hfil$\crcr#2\crcr\sim\crcr}}}
\def\la{\mathrel{\mathpalette\fun <}}
\def\ga{\mathrel{\mathpalette\fun >}}
\def\order{{\cal O}}

\def\sbar#1{\kern 0.8pt
        \overline{\kern -0.8pt #1 \kern -0.8pt}
        \kern 0.8pt}  

\def\meter{\ifmmode \,\, {\rm m} \else m \fi}
\def\yr {\ifmmode \,\, {\rm yr} \else yr \fi}

\def\sr{\ifmmode \,\, {\rm sr} \else sr \fi}

\def\hatn{{\bf \hat n}}

\def\slashchar#1{\setbox0=\hbox{$#1$}           
   \dimen0=\wd0                                 
   \setbox1=\hbox{/} \dimen1=\wd1               
   \ifdim\dimen0>\dimen1                        
      \rlap{\hbox to \dimen0{\hfil/\hfil}}      
      #1                                        
   \else					
      \rlap{\hbox to \dimen1{\hfil$#1$\hfil}}   
      /                                         
   \fi}

\begin{document}
\rightline{CU-TP-917, CAL-669, astro-ph/9809214~\footnote{To
appear in {\it The Early and Future Universe}, proceedings of
the CCAST workshop, Beijing, China, June 22--27, 1998, edited by
Minghan Ye (Gordon and Breach Publishers).}}
\vspace*{1cm}
\title{POSSIBLE RELICS FROM NEW PHYSICS IN THE EARLY UNIVERSE:
INFLATION, THE COSMIC MICROWAVE BACKGROUND, AND PARTICLE DARK
MATTER}

\author{MARC KAMIONKOWSKI~\footnote{kamion@phys.columbia.edu}}

\address{Department of Physics, Columbia University, 538 West
120th St., New York, NY 10027~~U.S.A}

\maketitle\abstracts{
I review two different connections between particle theory and
early-Universe cosmology: (1) Cosmic-microwave-background (CMB)
tests of inflation and (2) particle dark matter.  The
inflationary predictions of a flat Universe and a nearly
scale-invariant spectrum of primordial density perturbations
will be tested precisely with forthcoming maps of the CMB
temperature.  A stochastic gravitational-wave background may be
probed with a map of the CMB polarization.  I also discuss some
other uses of CMB maps.  Particle theory has produced two very
well-motivated candidates for the dark matter in the Universe:
an axion and a supersymmetric particle.  In both cases, there
are a variety of experiments afoot to detect these particles.  I 
review the properties of these dark-matter candidates and these
detection techniques.
}

\section{Introduction}

In recent decades, particle theorists and cosmologists have joined 
forces in an effort to uncover evidence for new physics
beyond the standard model while simultaneously trying to
reconstruct the events that occurred during the first second
after the big bang.  At first, this endeavor yielded a plethora
of ideas.  Although many of these early hypotheses have fallen
by the wayside, several have remained intact, become increasingly
attractive, and are currently under experimental scrutiny.  The
purpose of these lectures is to review what I believe to be two
of the currently most compelling connections between particle
physics and early-Universe cosmology: Cosmic microwave
background (CMB) tests of inflation and the particle solution to
the dark-matter problem.

\subsection{Inflation and the Cosmic Microwave Background}

Despite its major triumphs (the expansion, nucleosynthesis,
and the cosmic microwave background), the big-bang theory for
the origin of the Universe leaves several questions unanswered.
Chief amongst these is the horizon problem:  When cosmic
microwave background (CMB) photons last scattered, the age of
the Universe was roughly 100,000 years, much smaller than its
current age of roughly 10 billion years.  After taking into
account the expansion of the Universe, one finds that the angle
subtended by a causally connected region at the surface of last
scatter is roughly $1^\circ$. However, there are 40,000 square
degrees on the surface of the sky.  Therefore, when we look at
the CMB over the entire sky, we are looking at 40,000
causally disconnected regions of the Universe.  But quite remarkably,
each has the same temperature to roughly one part in $10^5$!

The most satisfying (only?) explanation for this is slow-roll
inflation \cite{inflation}, a
period of accelerated expansion in the early Universe driven by
the vacuum energy most likely associated with a symmetric phase
of a GUT Higgs field (or perhaps Planck-scale physics or
Peccei-Quinn symmetry breaking).  Although the physics
responsible for inflation is still not well understood,
inflation generically predicts (1) a flat Universe; (2) that
primordial adiabatic (i.e., curvature) perturbations are
responsible for the large-scale structure (LSS) in the Universe
today \cite{scalars}; and (3) a stochastic gravity-wave
background \cite{abbott}.  More
precisely, inflation predicts a spectrum $P_s = A_s k^{n_s}$
(with $n_s$ near unity) of primordial density (scalar metric)
perturbations, and a stochastic gravity-wave background (tensor
metric perturbations) with spectrum $P_t =A_t \propto k^{n_t}$
(with $n_t$ small compared with unity).  (4) Inflation further
uniquely predicts specific relations between the
``inflationary observables,'' the amplitudes $A_s$ and $A_t$ and
spectral indices $n_s$ and $n_t$ of the scalar and tensor
perturbations \cite{steinhardt}. The amplitude of the
gravity-wave background is
proportional to the height of the inflaton potential, and the
spectral indices depend on the shape of the inflaton potential.
Therefore, determination of these parameters would illuminate
the physics responsible for inflation.  

Until recently, none of these predictions could really be tested.
Measured values for the density of the Universe span
almost an order of magnitude.  Furthermore, most do not probe
the possible contribution of a cosmological constant (or some
other diffuse matter component), so they do not address the
geometry of the Universe.  The only observable effects of a
stochastic gravity-wave background are in the CMB.  COBE
observations do in fact provide an upper limit to the tensor
amplitude, and therefore an inflaton-potential height near the
GUT scale.  However, there is no way to disentangle the scalar
and tensor contributions to the COBE anisotropy.

It has recently become increasingly likely that
adiabatic perturbations are responsible for the
origin of structure.  Before COBE, there were numerous plausible
models for structure formation: e.g., isocurvature perturbations
both with and without cold dark matter, late-time or slow phase 
transitions, topological defects (cosmic strings or textures),
superconducting cosmic strings, explosive or seed models, a
``loitering'' Universe, etc.  However, the amplitude of the COBE 
anisotropy makes all these alternative models unlikely.  With
adiabatic perturbations, hotter regions at the surface of last
scatter are embedded in deeper potential wells, so the
reddening due to the the gravitational redshift of the photons
from these regions partially cancels the higher intrinsic
temperatures.  Thus, other models will generically produce more
anisotropy for the same density perturbation.  When
normalized to the density fluctuations indicated by galaxy
surveys, alternative models thus generically produce a larger
temperature fluctuation than that measured by COBE \cite{jaffe}.
In the past year, some leading proponents of topological
defects, the leading alternative, have conceded that these
models have difficulty accounting for the origin of large-scale
structure \cite{towel}.

We are now entering an exciting new era, driven by new
theoretical ideas and developments in detector technology, in
which the predictions of inflation will be tested with
unprecedented precision.  It is even conceivable that early in
the next century, we will move from verification to
direct investigation of the high-energy physics responsible for
inflation.

\subsection{Dark Matter and New Particles}

Another of the grand cosmic mysteries today is the nature of the 
dark matter.  Almost all astronomers will agree that most of the 
mass in the Universe is nonluminous.  Dynamics of clusters of
galaxies suggest a universal nonrelativistic-matter density of
$\Omega_0\simeq0.1-0.3$.  If the luminous matter were all there
was, the duration of the epoch of structure formation would be
very short, thereby requiring (in almost all theories of
structure formation) fluctuations in the CMB which would be
larger than those observed.  These considerations imply
$\Omega_0\ga0.3$ \cite{kamspergelsug}.  Second, if the current
value of $\Omega_0$ is of order unity today, then at the Planck
time it must have been $1\pm10^{-60}$ leading us to believe that
$\Omega_0$ is precisely unity for aesthetic reasons.  And of
course, inflation, must set $\Omega$, the {\it total} density
(including a cosmological constant), to unity.

\begin{figure}[htbp]
\centerline{\psfig{file=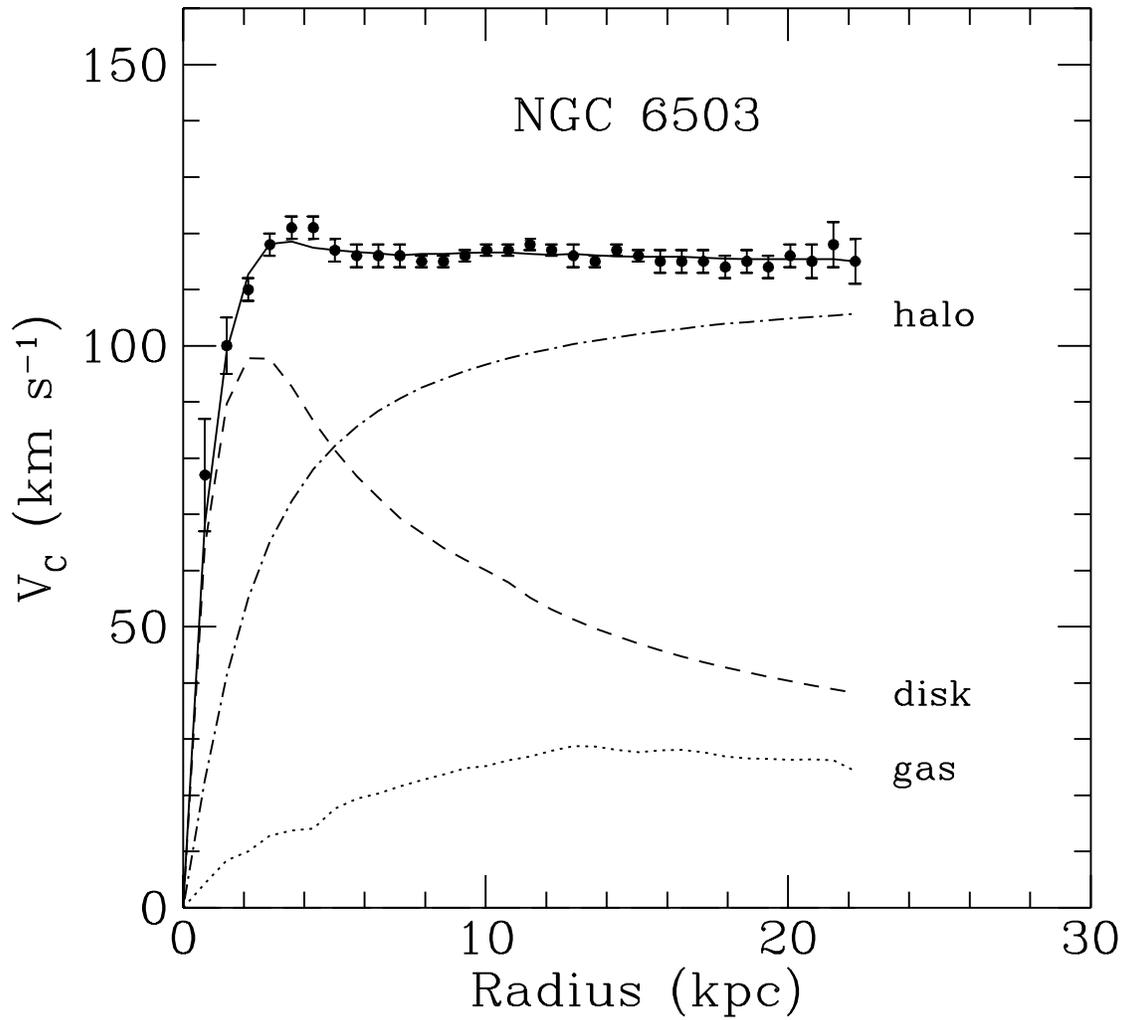,width=6in}}
\caption{Rotation curve for the spiral galaxy NGC6503.  The points
	 are the measured circular rotation velocities as a
	 function of distance from the center of the galaxy.
	 The dashed and dotted curves are the contribution to
	 the rotational velocity due to the observed disk and
	 gas, respectively, and the dot-dash curve is the
	 contribution from the dark halo.}
\label{rotationfigure}
\end{figure}

However, the most robust observational evidence for the
existence of dark matter involves galactic dynamics.  There is
simply not enough luminous matter ($\Omega_{\rm lum}\la0.01$)
observed in spiral galaxies to account for their observed
rotation curves (for example, that for NGC6503 shown in
Fig.~\ref{rotationfigure} \cite{broeils}).  Newton's
laws imply galactic dark halos with masses that contribute
$\Omega_{\rm halo} \ga 0.1$.

On the other hand, big-bang nucleosynthesis
suggests that the baryon density is $\Omega_b\la0.1$ \cite{bbn}, too
small to account for the dark matter in the Universe.
Although a neutrino species of mass ${\cal O}(30\, {\rm eV})$ could
provide the right dark-matter density, N-body simulations of
structure formation in a neutrino-dominated Universe do a poor
job of reproducing the observed structure \cite{Nbody}.
Furthermore, it is difficult to see (essentially from the Pauli
principle) how such a neutrino could make up the dark matter in
the halos of galaxies \cite{gunn}. It appears likely then, that some
nonbaryonic, nonrelativistic matter is required.

The two leading candidates from particle theory are the
axion \cite{axion}, which arises in the Peccei-Quinn solution to
the strong-$CP$ problem, and a weakly-interacting massive particle
(WIMP), which may arise in supersymmetric (or other) extensions
of the standard model \cite{jkg}.  As discussed below, there are
good reasons to believe that if the Peccei-Quinn mechanism is
responsible for preserving $CP$ in the strong
interactions, then the axion is the dark matter.  Similarly,
there are also excellent reasons to expect that if low-energy
supersymmetry exists in Nature, then the dark matter should be
composed of the lightest superpartner.  The study of these ideas
are no longer exclusively the domain of theorists: there are now
a number of experiments aimed at discovery of axions and WIMPs.
If axions populate the Galactic halo, they can be converted to
photons in resonant cavities immersed in strong magnetic
fields.  An experiment to search for axions in this fashion is
currently being carried out.  If WIMPs populate the halo, they
can be detected either directly in low-background laboratory
detectors or indirectly via observation of energetic neutrinos
from WIMPs which have accumulated and then annihilated in the
Sun and/or Earth.

\subsection{Outline}

One of the purposes of these lectures is to review how
forthcoming CMB experiments will test several of the predictions
of inflation.  I will not have time to provide a real
review of inflation.  For more details on inflation (as well as
the standard cosmological model), see the books by Kolb
and Turner \cite{kolbturnerbook} and/or Peebles \cite{peebles},
or the lectures in these proceedings by Steinhardt.
Here, I will first review the predictions of inflation for density
perturbations and gravity waves.  I will then discuss how CMB
temperature anisotropies will test the inflationary predictions
of a flat Universe and a primordial spectrum of density
perturbations.  I review how a CMB polarization map
may be used to isolate the gravity waves and briefly review how
detection of these tensor modes may be used to learn about the
physics responsible for inflation.  

Although my focus here is on CMB tests of inflation, the CMB may 
also be useful for addressing some other issues in cosmology
that may be of only tangential relevance to inflation, and I
briefly review a few of those here.  I discuss some recent
work on secondary anisotropies at smaller angular scales, and
how these may be used to probe the epoch at which objects first
underwent gravitational collapse in the Universe.  I then
briefly mention a few other possibilities.

In the Section on dark matter, I first show how the observed
dynamics of the Milky Way indicate a local dark-matter density
of $\rho_0\simeq0.4$  GeV~cm$^{-3}$.  I then review the
arguments for axion and WIMP dark matter and the methods of
detection.  However, there is no way I can do this very active
field of research justice in such a short space.  Readers with further
interest in WIMPs should see the review article by
Jungman, Griest and me \cite{jkg}.  The first four Sections of
that article are meant to provide a general review of dark and
supersymmetry, and  the idea of WIMP dark matter.  The remainder
of that article provides technical details required by those
interested in actively pursuing research on the topic.  There
are also several excellent axion reviews \cite{axion} and the
recent book by Raffelt \cite{raffelt}.

\section{The Cosmic Microwave Background and Inflation}

\subsection{Inflationary Observables}

Inflation occurs when the energy density of the Universe is
dominated by the vacuum energy $V(\phi)$ associated with some
scalar field $\phi$ (the ``inflaton'').  During this time, the
quantum fluctuations in $\phi$ produce classical scalar
perturbations, and quantum fluctuations in the spacetime metric
produce gravitational waves.  If the inflaton potential
$V(\phi)$ is given in units of $m_{\rm Pl}^4$, and the inflaton
$\phi$ is in units of $m_{\rm Pl}$, then the scalar and tensor
spectral indices are
\begin{eqnarray}
     1-n_s &=& { 1 \over 8\pi} \left( {V' \over V} \right)^2 -
     {1 \over 4 \pi} \left({V' \over V} \right)', \nonumber\\
     n_t &=& -{ 1 \over 8\pi} \left( {V' \over V} \right)^2. 
\label{spectralindices}
\end{eqnarray}
The amplitudes can be fixed by their contribution to $C_2^{\rm TT}$,
the quadrupole moment of the CMB temperature,
\begin{eqnarray}
     {\cal S} &\equiv & 6\, C_2^{{\rm TT},{\rm scalar}}=
     33.2\,[V^3/(V')^2],
          \nonumber\\
     {\cal T} &\equiv & 6\, C_2^{{\rm TT},{\rm tensor}}= 9.2 \,V.
\label{amplitudes}
\end{eqnarray}
For the slow-roll conditions to be satisfied, we must have
\begin{eqnarray}
     (1 /16 \pi) (V'/V)^2 &\ll& 1, \\ 
     (1 /8\pi)(V''/V) & \ll & 1,
\label{slowrollconditions}
\end{eqnarray}
which guarantee that inflation lasts long enough to make the Universe
flat and to solve the horizon problem.

When combined with COBE results, current degree-scale--anisotropy and
large-scale-structure observations suggest that ${\cal T}/{\cal S}$ is less
than order unity in inflationary models, which restricts
$V\la 5\times 10^{-12}$.  Barring strange coincidences, the COBE
spectral index and relations above suggest that if slow-roll
inflation is right, then the scalar and tensor spectra must both be
nearly scale invariant ($n_s\simeq 1$ and $n_t\simeq 0$).

\subsection{Temperature Anisotropies}

The primary goal of CMB experiments that map the temperature as
a function of position on the sky is recovery of the
temperature autocorrelation function or angular power spectrum
of the CMB.  The fractional temperature perturbation
$\Delta T(\hatn)/T$ in a given direction $\hatn$ can be expanded
in spherical harmonics,
\begin{equation}
     {\Delta T(\hatn) \over T} = \sum_{lm} \, a_{(lm)}^{\rm T}\,
     Y_{(lm)}(\hatn), \quad {\rm with} \quad     a_{(lm)}^{\rm
     T} = \int\, d\hatn\, Y_{(lm)}^*(\hatn) \, {\Delta T(\hatn)
     \over T}.
\label{eq:Texpansion}
\end{equation}
Statistical isotropy and homogeneity of the Universe imply that
these coefficients have expectation values $\VEV{ (a_{(lm)}^{\rm
T})^*
a_{(l'm')}^{\rm T}} = C_l^{\rm TT} \delta_{ll'} \delta_{mm'}$ when
averaged over the sky.  Roughly speaking, the multipole moments
$C_l^{\rm TT}$ measure the mean-square temperature difference
between two points separated by an angle $(\theta/1^\circ) \sim
200/l$.

\begin{figure}[htbp]
\centerline{\psfig{file=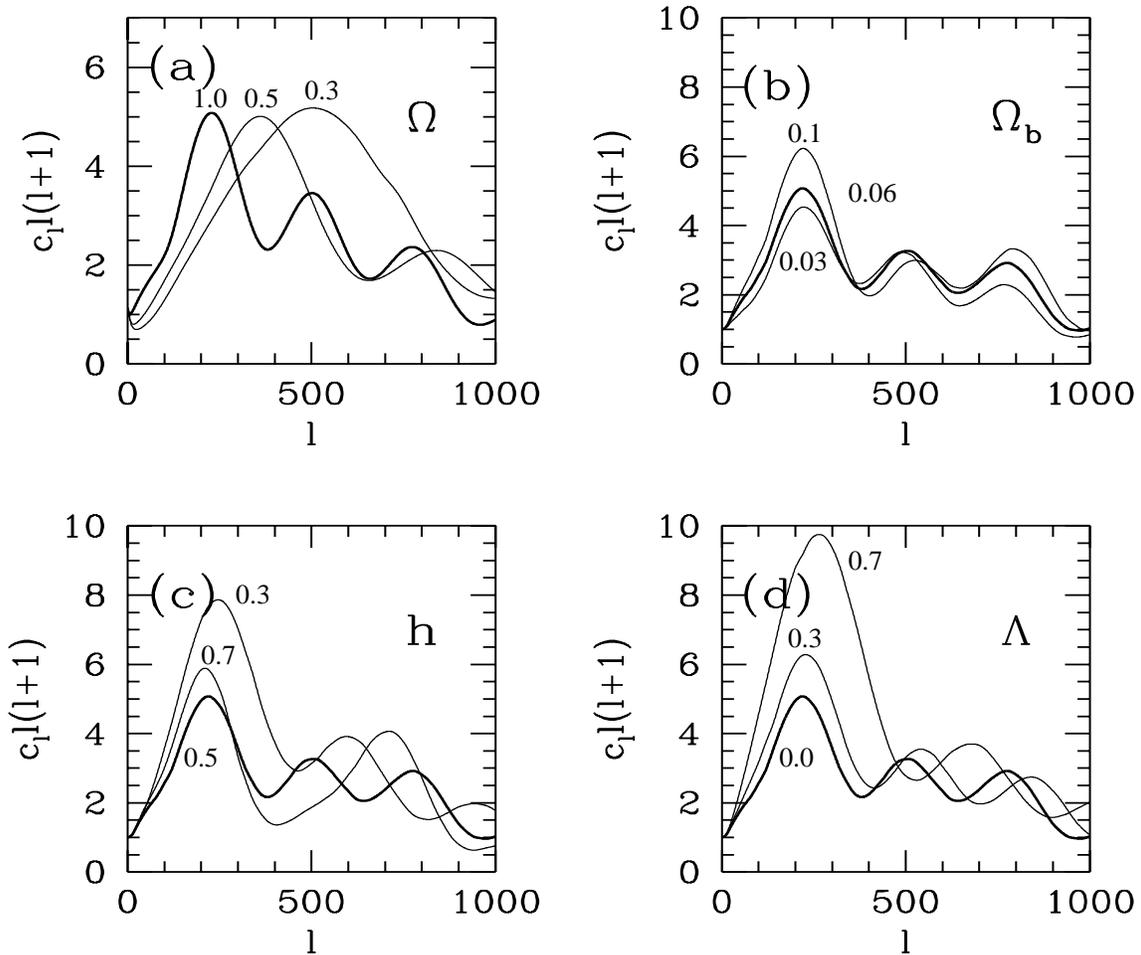,width=6in}}
\caption{
	  Theoretical predictions for CMB spectra as a function
	  of multipole moment $l$ for models with primordial
	  adiabatic perturbations.  In each case, the
	  heavy curve is that for the canonical standard-CDM values,
	  a total density $\Omega=1$, cosmological constant
	  $\Lambda=0$, baryon density $\Omega_b=0.06$, and
	  Hubble parameter $h=0.5$.  Each graph shows the effect
	  of variation of one of these parameters.  In (d),
	  $\Omega_0+\Lambda=1$.
}
\label{fig:models}
\end{figure}

Predictions for the $C_l$'s can be made given a theory for
structure formation and the values of several cosmological
parameters: the total density $\Omega$ in units of the critical
density, the cosmological-constant $\Lambda$ in units of the
critical density, the Hubble constant $h$ in units of
100~km~sec$^{-1}$~Mpc$^{-1}$, and the baryon density $\Omega_b$
in units of the critical density.  Fig.~\ref{fig:models} shows
predictions for
models with primordial adiabatic perturbations.  The wriggles
come from oscillations in the photon-baryon fluid at the surface
of last scatter.  Each panel shows the effect of independent
variation of one of the cosmological parameters.  As
illustrated, the height, width, and spacing of the acoustic
peaks in the angular spectrum depend on  these (and other)
cosmological parameters.

These small-angle CMB anisotropies can be used to determine the
geometry of the Universe \cite{kamspergelsug}.  The angle
subtended by the horizon at the surface of last scatter is
$\theta_H \sim \Omega^{1/2} \;1^\circ$, and the peaks in the CMB
spectrum are due to causal processes at the surface of last
scatter.  Therefore, the angles (or values of $l$) at which the
peaks occur determine the geometry of the Universe.  This is
illustrated in Fig.~\ref{fig:models}(a) where the CMB spectra
for several values of $\Omega$ are shown.  As illustrated in the
other panels, the angular position of the first  peak is
relatively insensitive to the values of other undetermined (or
still imprecisely determined) cosmological parameters such as
the baryon density, the Hubble constant, and the cosmological
constant (as well as several others not shown such as the
spectral indices and amplitudes of the scalar and tensor spectra
and the ionization history of the Universe).  Therefore,
determination of the location of this first acoustic peak should
provide a robust measure of the geometry of the Universe.

\begin{figure}[htbp]
\centerline{\psfig{file=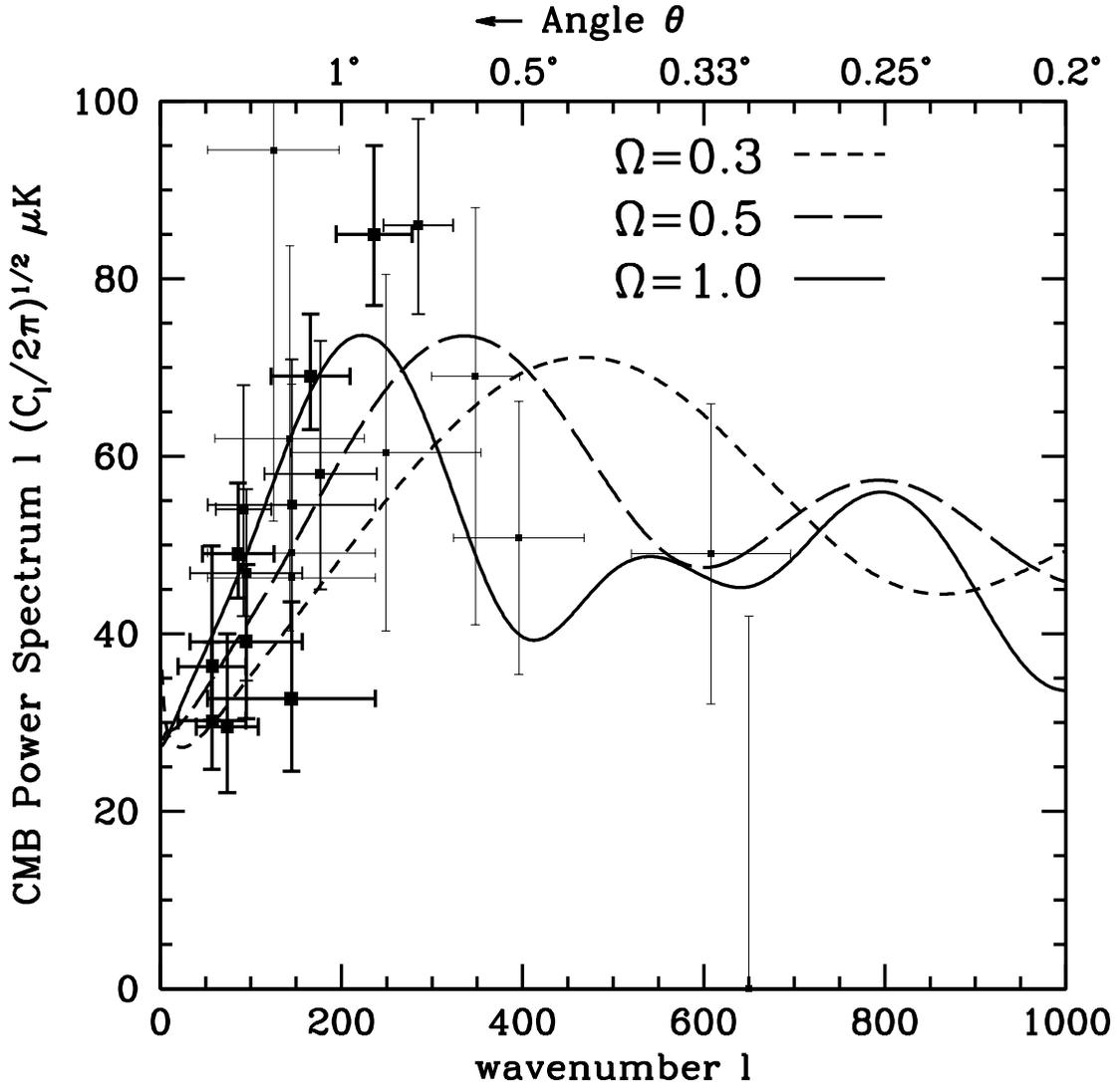,width=6in}}
\caption{Current CMB data (from Ref. \protect\cite{science}).}
\label{fig:data}
\end{figure}

Fig.~\ref{fig:data} shows data from current ground-based and
balloon-borne experiments.  By fitting the theoretical curves to
these points, several groups find that the best fit to the data
is found with a total density $\Omega\simeq1.0$ \cite{current}.
However, visual inspection of the data points in
Fig.~\ref{fig:data} clearly indicate that this current
determination of the geometry cannot be robust.

\begin{figure}[htbp]
\centerline{\psfig{file=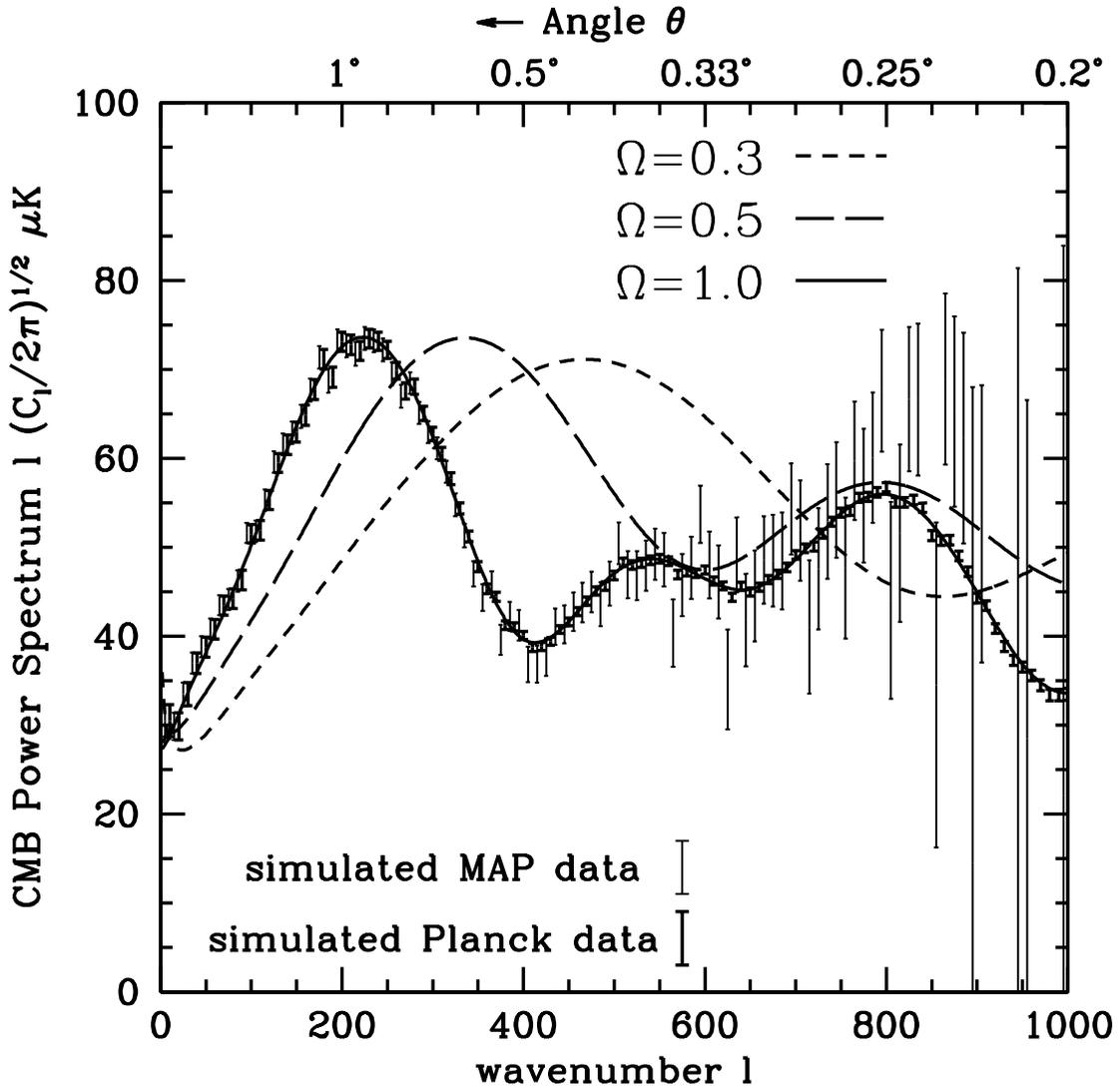,width=6in}}
\caption{Simulated MAP and Planck data (from Ref. \protect\cite{science}).}
\label{fig:simulated}
\end{figure}

In the near future, the precision with which this determination
can be made will be improved dramatically.  NASA has recently
approved the flight of a satellite mission, the Microwave
Anisotropy Probe (MAP) \cite{MAP} in the year 2000 to carry out
these measurements, and the European Space Agency (ESA) has
approved the flight of a subsequent more precise experiment, the Planck
Surveyor \cite{PLANCK}.  Fig.~\ref{fig:simulated} shows simulated
data from MAP and Planck.  The heavier points with smaller error bars
are those we might expect from Planck and the lighter points
with larger error bars are those anticipated for MAP.  Even
without any sophisticated analysis, it is clear from
Fig.~\ref{fig:simulated} that data from either of these
experiments will be able to locate the first acoustic peak
sufficiently well to discriminate between a flat Universe
($\Omega=1$) and an open Universe with $\Omega\simeq0.3-0.5$.

By doing what essentially boils down to a calculation of the
covariance matrix for such simulated data,
it can be shown that these satellite missions may
potentially determine $\Omega$ to better than 10\% {\it after}
marginalizing over all other undetermined parameters (we
considered 7 more parameters in addition to the 4 shown in
Fig.~\ref{fig:models}), and better than 1\% if the other parameters can be
fixed by independent observations or assumption \cite{jkksone}.
This would be far more accurate than any traditional
determinations of the geometry.

It can similarly be shown that the CMB should provide values for
the cosmological constant and baryon density far more precise
than those from traditional observations \cite{jkkstwo,bet}.  If there is more
nonrelativistic matter in the Universe than baryons can account
for---as suggested by current observations---it will become
increasingly clear with future CMB measurements.  

Although these initial forecasts relied on the assumptions that
adiabatic perturbations were responsible for structure formation
and that reionization would not erase CMB anisotropies, these
assumptions have become increasingly
justifiable in the past few years.  As discussed above,
the leading alternative theories for structure formation now
appear to be in trouble, and recent detections of CMB
anisotropy at degree angular separations show that the effects
of reionization are small.  

The predictions of a nearly scale-free spectrum of primordial
adiabatic perturbations will also be further tested with
measurements of small-angle CMB anisotropies.  The existence and
structure of the acoustic peaks shown in Fig.~\ref{fig:models}
will provide an unmistakable signature of adiabatic
perturbations \cite{huwhite} and the spectral index $n_s$ can be determined
from fitting the theoretical curves to the data in the same way
that the density, cosmological constant, baryon density, and
Hubble constant are also fit \cite{jkkstwo}.

Temperature anisotropies produced by a stochastic gravity-wave
background would affect the shape of the angular CMB spectrum,
but there is no way to disentangle the scalar and tensor
contributions to the CMB anisotropy in a model-independent way.
Unless the tensor signal is large, the cosmic variance from the
dominant scalar modes will provide an irreducible limit to the
sensitivity of a temperature map to a tensor signal \cite{jkkstwo}.

\subsection{CMB Polarization and Gravitational Waves}

Although a CMB temperature map cannot unambiguously distinguish
between the density-perturbation and gravity-wave contributions
to the CMB, the two can be decomposed in a model-independent
fashion with a map of the CMB
polarization \cite{probe,ourpolarization,selzald}.  Suppose we 
measure the linear-polarization ``vector'' $\vec P(\hatn)$ at
every point $\hatn$ on the sky.  Such a vector field can be written as the
gradient of a scalar function $A$ plus the curl of a vector
field $\vec B$: $\vec P(\hatn) \, = \, \vec \nabla A \, + \,
\vec\nabla \times \vec B.$
The gradient (i.e., curl-free) and curl components can be
decomposed by taking the divergence or curl of $\vec
P(\hatn)$ respectively.  Density perturbations are scalar metric
perturbations, so they have no handedness.  They can therefore
produce no curl.  On the other hand, gravitational waves {\it
do} have a handedness so they can (and we have shown that they
do) produce a curl.  This therefore provides a way to detect the
inflationary stochastic gravity-wave background and thereby
test the relations between the inflationary observables.  It
should also allow one to determine (or at least constrain in the
case of a nondetection) the height of the inflaton potential.

More precisely, the Stokes parameters $Q(\hatn)$ and $U(\hatn)$
(where $Q$ and $U$ are measured with respect to the polar ${\bf
\hat\theta}$ and azimuthal ${\bf \hat \phi}$ axes) which specify
the linear polarization in direction $\hatn$ are components of a
$2\times2$ symmetric trace-free (STF) tensor, 
\begin{equation}
  {\cal P}_{ab}(\hatn)={1\over 2} \left( \begin{array}{cc}
   \vphantom{1\over 2}Q(\hatn) & -U(\hatn) \sin\theta \\
   -U(\hatn)\sin\theta & -Q(\hatn)\sin^2\theta \\
   \end{array} \right),
\label{whatPis}
\end{equation}
where the subscripts $ab$ are tensor indices.
Just as the temperature is expanded in terms of spherical
harmonics, the polarization tensor can be expanded \cite{ourpolarization},
\begin{equation}
      {{\cal P}_{ab}(\hatn)\over T_0} =
      \sum_{lm} \Biggl[ a_{(lm)}^{{\rm G}}Y_{(lm)ab}^{{\rm
      G}}(\hatn) +a_{(lm)}^{{\rm C}}Y_{(lm)ab}^{{\rm C}}(\hatn)
      \Biggr],
\label{Pexpansion}
\end{equation}
in terms of the tensor spherical harmonics $Y_{(lm)ab}^{\rm G}$
and $Y_{(lm)ab}^{\rm C}$, which are a complete basis for the
``gradient'' (i.e., curl-free) and ``curl'' components of the
tensor field, respectively.  The mode amplitudes are given by
\begin{equation}
a^{\rm G}_{(lm)}={1\over T_0}\int d\hatn\,{\cal P}_{ab}(\hatn)\, 
                                         Y_{(lm)}^{{\rm G}
					 \,ab\, *}(\hatn),
\qquad a^{\rm C}_{(lm)}={1\over T_0}\int d\hatn\,{\cal P}_{ab}(\hatn)\,
                                          Y_{(lm)}^{{\rm C} \,
					  ab\, *}(\hatn).
\label{Amplitudes}
\end{equation}
Here $T_0$ is the cosmological mean CMB temperature and $Q$ and
$U$ are given in brightness temperature units rather than flux
units.   Scalar perturbations have no handedness.  Therefore,
they can produce no curl, so $a_{(lm)}^{\rm C}=0$ for scalar
modes.  On the other hand tensor modes {\it do} have a
handedness, so they produce a non-zero curl, $a_{(lm)}^{\rm C}
\neq0$.

A given inflationary model predicts that the $a_{(lm)}^{\rm X}$
are gaussian random variables with zero mean,
$\VEV{a_{(lm)}^{\rm X}}=0$  (for ${\rm X},{\rm X}' = \{{\rm
T,G,C}\}$) and covariance $\VEV{\left(a_{(l'm')}^{\rm X'}
\right)^* a_{(lm)}^{\rm X}} = C_l^{{\rm XX}'}
\delta_{ll'}\delta_{mm'}$.   Parity demands that
$C_l^{\rm TC}=C_l^{\rm GC}=0$.  Therefore the statistics of the
CMB temperature-polarization map are completely specified by the
four sets of moments, $C_l^{\rm TT}$, $C_l^{\rm TG}$, $C_l^{\rm
GG}$, and $C_l^{\rm CC}$.   Also, as stated above, only tensor modes
will produce nonzero $C_l^{\rm CC}$.  

\begin{figure}[htbp]
\centerline{\psfig{file=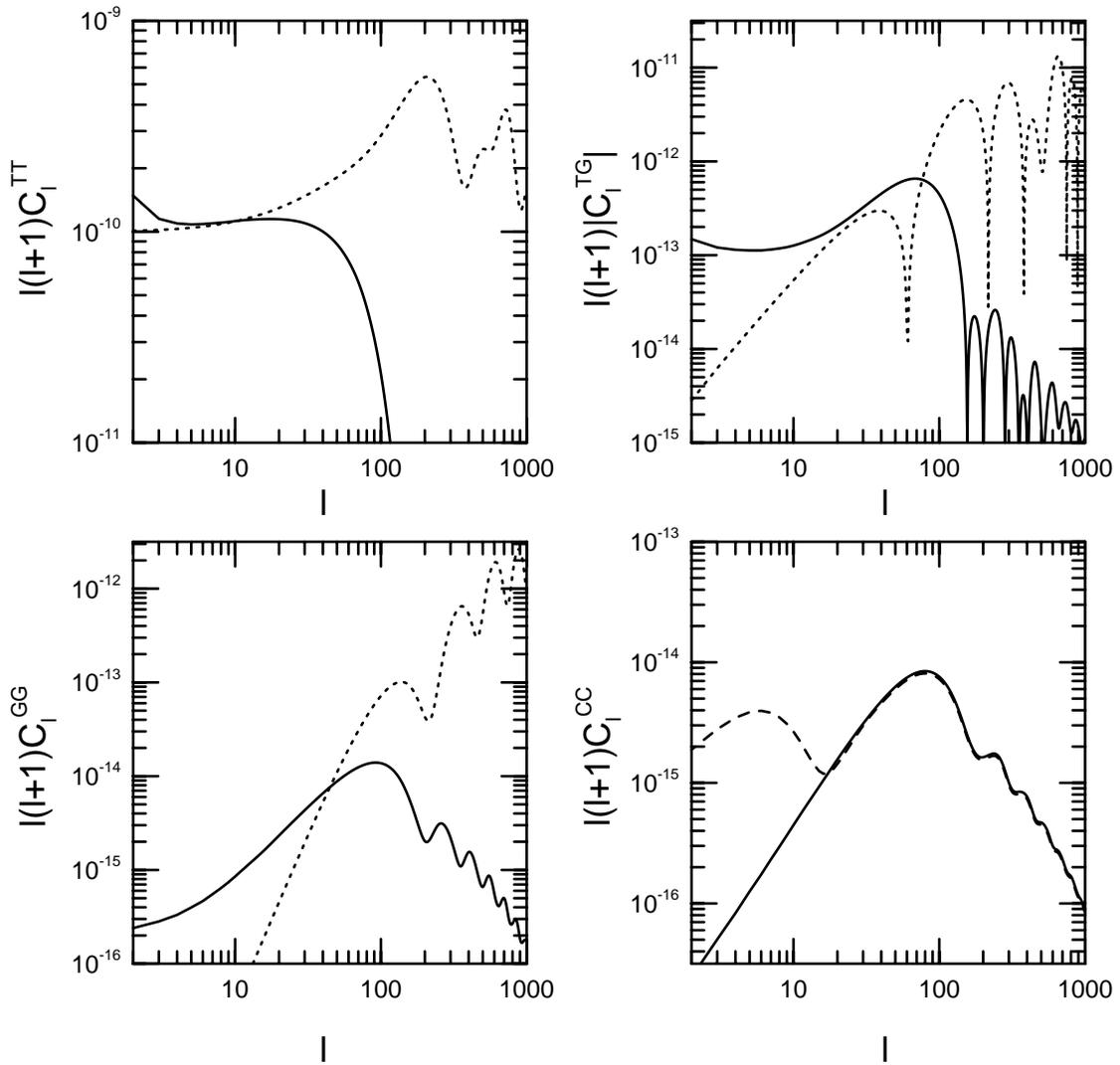,width=8in}}
\caption{
          Theoretical predictions for the four nonzero CMB
	  temperature-polarization spectra as a function
	  of multipole moment $l$.  The dashed line in the lower
	  right panel shows a reionized model with optical depth
	  $\tau=0.1$ to the surface of last scatter.
       }
\label{clsplot}
\end{figure}

To~illustrate, Fig.~\ref{clsplot} shows the four
temperature-polarization power spectra. The dotted curves
correspond to a COBE-normalized inflationary model
with cold dark matter and no cosmological constant
($\Lambda=0$), Hubble constant (in units of 100
km~sec$^{-1}$~Mpc$^{-1}$) $h=0.65$, baryon density
$\Omega_bh^2=0.024$, scalar spectral index $n_s=1$, no
reionization, and no gravitational waves.  The solid curves show
the spectra for a COBE-normalized stochastic gravity-wave
background with a flat scale-invariant spectrum ($h=0.65$,
$\Omega_b h^2=0.024$, and $\Lambda=0$) in a critical-density
Universe.   Note that the panel for $C_l^{\rm CC}$ contains no
dotted curve since scalar perturbations produce no C
polarization component.  The dashed curve in the CC panel shows
the tensor spectrum for a reionized model with optical depth
$\tau=0.1$ to the surface of last scatter.

As with a temperature map, the sensitivity of a polarization map
to gravity waves will be determined by the
instrumental noise and fraction of sky covered, and by the
angular resolution.  Suppose the detector sensitivity is $s$ and
the experiment lasts for $t_{\rm yr}$ years with an angular
resolution better than $1^\circ$.  Suppose further that we
consider only the CC component of the polarization in our
analysis.  Then the smallest tensor amplitude ${\cal T}_{\rm
min}$ to which the experiment will be sensitive at $1\sigma$
is \cite{detectability}
\begin{equation}
     {{\cal T}_{\rm min}\over 6\, C_2^{\rm TT}}
      \simeq 5\times 10^{-4} \left( {s\over \mu{\rm K}\,\sqrt{\rm
      sec}} \right)^2 t_{\rm yr}^{-1}.
\label{CCresult}
\end{equation}
Thus, the curl component of a full-sky polarization map is
sensitive to inflaton potentials $V\ga 5 \times
10^{-15}t_{\rm yr}^{-1}$ $(s/\mu{\rm K}\, \sqrt{\rm sec})^2$.  
Improvement on current constraints with only the curl
polarization component requires a detector sensitivity
$s\la40\,t_{\rm yr}^{1/2}\,\mu$K$\sqrt{\rm sec}$.  For
comparison, the detector sensitivity of MAP will be $s={\cal
O}(100\,\mu$K$\sqrt{\rm sec})$.  However, Planck may conceivably
get sensitivities around $s=25\,\mu$K$\sqrt{\rm sec}$.

\begin{figure}[htbp]
\centerline{\psfig{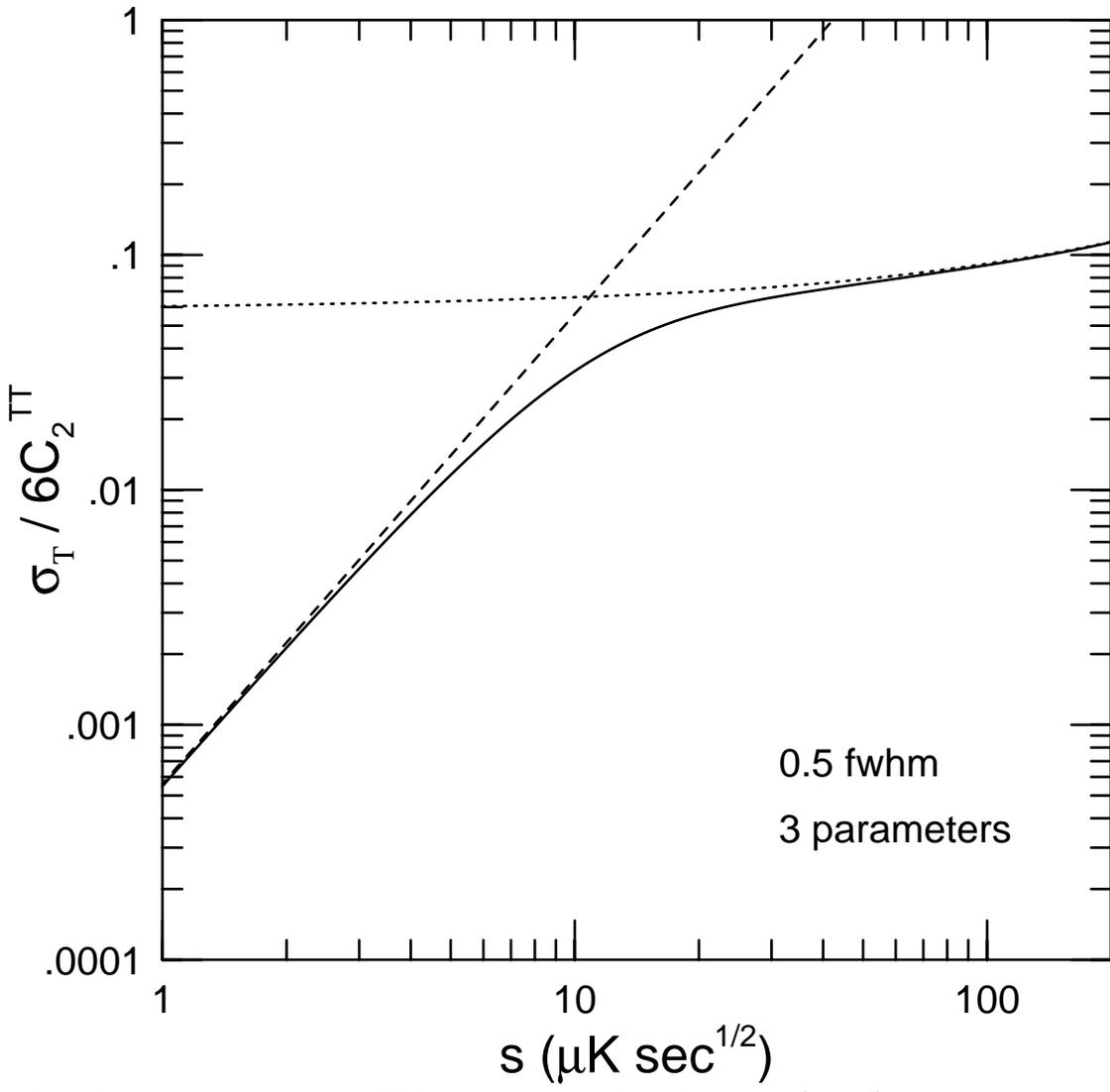}}
\caption{
         Results for the $1\sigma$ sensitivity $\sigma_{\cal T}$ to
	 the amplitude $\cal T$ of a flat ($n_t=0$) tensor spectrum
	 as a function of detector sensitivity $s$ for an
	 experiment which maps the CMB temperature and
	 polarization on the full sky for one year with an
	 angular resolution of $0.5^\circ$.  The vertical axis
	 is in units of the temperature quadrupole.  See text
	 for more details.}
\label{resultssimple}
\end{figure}

Eq.~(\ref{CCresult}) is the sensitivity obtained by using only
the curl component of the polarization, which provides a
model-independent probe of the tensor signal.  However, if we
are willing to consider specific models for the tensor and
scalar spectra, the sensitivity to a tensor signal may be
improved somewhat by considering the predictions for the full
temperature/polarization auto- and cross-correlation power
spectra \cite{detectability}.

For example, Fig.~\ref{resultssimple} shows the $1\sigma$
sensitivity $\sigma_{\cal T}$ to the amplitude $\cal T$ of a
flat ($n_t=0$) tensor spectrum as a function of detector
sensitivity $s$ for an experiment which maps the CMB temperature
and polarization on the full sky for one year with an angular
resolution of $0.5^\circ$.  The dotted curve
shows the results obtained by fitting only the TT moments; the
dashed curve shows results obtained by fitting only the CC
moments; and the solid curve shows results obtained by fitting
all four nonzero sets of moments. 

In Fig.~\ref{resultssimple}, we have assumed that the
spectra are fit only to $\cal S$, $\cal T$, and $n_s$, and the
parameters of the cosmological model are those used in
Fig.~\ref{resultssimple}.  If one fits to more cosmological
parameters (e.g., $\Omega_b$, $h$, $\Lambda$, etc.) as well, the
the sensitivity from the temperature moments, important for
larger $s$, is degraded.  However, the sensitivity due to the CC
component, which controls the total sensitivity for smaller $s$,
is essentially unchanged.  Again, this is because the CC signal
is very model-independent.

For detectors sensitivities $s\ga 20\,\mu$K$\,
\sqrt{\rm sec}$, the tensor-mode detectability for the
three-parameter fit shown in Fig.~\ref{resultssimple} comes
primarily from the temperature map, although polarization does
provides some incremental improvement.  However, if the data are
fit to more cosmological parameters (not shown), the
polarization improves the tensor sensitivity even for $s\ga
20\,\mu$K$\, \sqrt{\rm sec}$.  In any case, the
sensitivity to tensor modes comes almost entirely from the
curl of the polarization for detector sensitivities $s
\la10\,\mu$K$\sqrt{\rm sec}$.  Since the value of $s$ for Planck
will be somewhat higher, it will likely require a more
sensitive future experiment to truly capitalize on the
model-independent curl signature of tensor modes.

One can also investigate how sensitively the inflationary
observables ($n_s$, $n_t$, and ${\cal T}/{\cal S}$) can be
determined by fitting the CMB power spectra measured, e.g., with 
MAP and Planck, with the CMB, both with and without a
polarization map.  One can then go one step further and see how
precisely the inflaton potential can be reconstructed from these 
measurements.  Ref. \cite{kinney} addresses these questions very
nicely.  One finds, that by fitting to all four nonzero 
sets of power spectra, inclusion of the polarization channels on
Planck will provide a dramatic improvement in constraints to
inflationary models.

Finally, it should be noted that even a small amount of
reionization will significantly increase the polarization signal
at low $l$ \cite{reionization} as shown in the CC panel of
Fig.~\ref{clsplot} for $\tau=0.1$.  With such a level of
reionization [which may be expected in inflation-inspired
cold-dark-matter (CDM) models, as discussed
in the following Section], the sensitivity to the tensor amplitude is
increased by more than a factor of 5 over that in
Eq.~(\ref{CCresult}).  This level of reionization (if not more)
is expected in cold dark matter
models \cite{kamspergelsug,blanchard,haiman}, so if anything,
Eq.~(\ref{CCresult}) and Fig.~\ref{resultssimple} provide
conservative estimates.

\subsection{Other Uses of the Cosmic Microwave Background}

{\it The Ostriker-Vishniac Effect and the Epoch of Reionization:} 
Although most of the matter in CDM models does not undergo
gravitational collapse until relatively late in the history of
the Universe, some small fraction of the mass is expected to
collapse at early times.  Ionizing radiation released by this
early generation of star and/or galaxy formation will partially
reionize the Universe, and these ionized electrons
will re-scatter at least some cosmic microwave background (CMB)
photons after recombination
at a redshift of $z\simeq1100$.  Theoretical
uncertainties in the process of star formation and the resulting
ionization make precise predictions of the ionization history
difficult.  Constraints to the shape of the CMB blackbody
spectrum and detection of CMB anisotropy at degree angular
scales suggest that if reionization occurred, the fraction of
CMB photons that re-scattered is small.  Still, estimates show
that even if small, at least some reionization is expected in
CDM models \cite{kamspergelsug,blanchard,haiman}: for
example, the most careful recent calculations suggest a
fraction $\tau_r\sim0.1$ of CMB photons were re-scattered \cite{haiman}.

Scattering of CMB photons from ionized
clouds will lead to anisotropies at arcminute separations below
the Silk-damping scale (the Ostriker-Vishniac
effect) \cite{ostriker,andrewmarc}.  These
anisotropies arise at higher order in perturbation theory and
are therefore not included in the usual Boltzmann calculations
of CMB anisotropy spectra.  The level of anisotropy is
expected to be small and it has so far eluded detection.  However,
these anisotropies may be observable with forthcoming CMB
interferometry experiments \cite{interferometers} that probe the
CMB power spectrum at arcminute scales.

\begin{figure}[htbp]
\centerline{\psfig{file=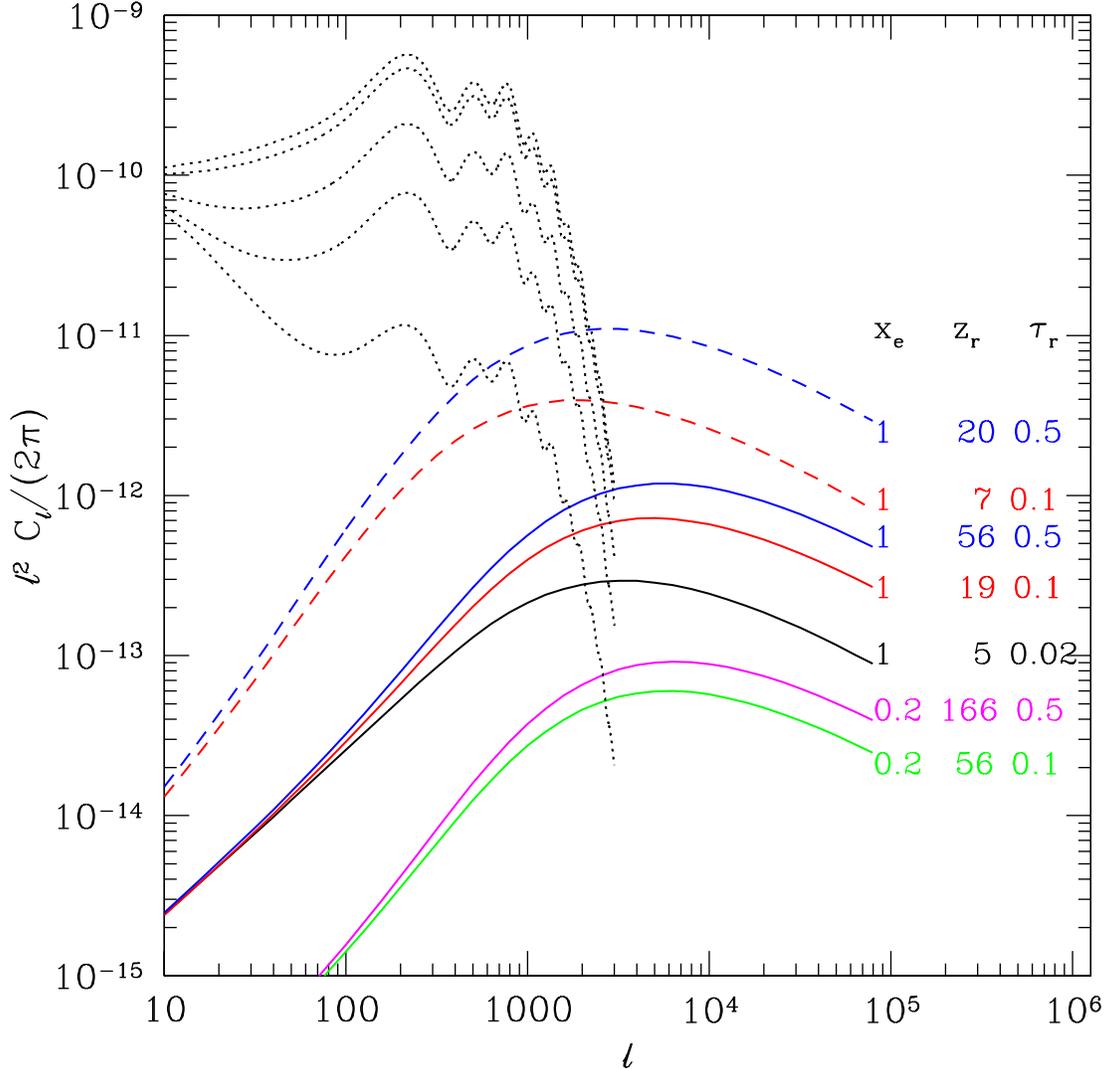,width=6in}}
\caption{Multipole moments for the Ostriker-Vishniac effect for
     the COBE-normalized canonical standard-CDM model
     ($\Omega=1$, $h=0.5$, $n=1$, $\Omega_b h^2 =0.0125$), for a
     variety of ionization histories.as listed. 
     We also show predictions for several open
     high-baryon-density models with the same $x_e$ and
     $\tau_r$, normalized to the cluster abundance, with dashed
     curves.  The dotted curves show the primary anisotropy for
     this model for $\tau_r=0.0$, 0.1, 0.5, 1, and 2, from top
     to bottom.
}
\label{fig:Clsplot}
\end{figure}

Fig.~\ref{fig:Clsplot} \cite{andrewmarc} shows the predicted
temperature-anisotropy spectrum from the Ostriker-Vishniac
effect for a number of ionization histories.  The
ionization histories are parameterized by an ionization fraction
$x_e$ and a redshift $z_r$ at which the Universe becomes
reionized.  The optical depth $\tau$ to the
standard-recombination surface of last scatter can be obtained
from these two parameters.  

Reionization damps the acoustic peaks in the
primary-anisotropy spectrum by $e^{-2\tau}$, as shown in
Fig.~\ref{fig:Clsplot}, but this damping is essentially
independent of the details of the ionization history.  That is,
any combination of $x_e$ and $z_r$ that gives the same $\tau$
has the same effect on the primary anisotropies.  So although
MAP and Planck will be able to determine $\tau$ from this
damping, they will not constrain the epoch of reionization.  On
the other hand, the secondary anisotropies (the
Ostriker-Vishniac anisotropies) produced at smaller angular
scales in reionized models {\it do} depend on the ionization
history.  For example, although the top and bottom dashed curves
in Fig.~\ref{fig:Clsplot} both have the same optical depth, they
have different reionization redshifts ($z_r=20$ and $z_r=57$).
Therefore, if MAP and Planck determine $\tau$, the amplitude of
the Ostriker-Vishniac anisotropy determines the reionization
epoch \cite{andrewmarc}.

In a flat Universe, CDM models  normalized to cluster abundances
produce rms temperature anisotropies of 0.8--2.4 $\mu$K on
arcminute angular scales for a constant ionization fraction of
unity, whereas an ionization fraction of 0.2 yields rms
anisotropies of 0.3--0.8 $\mu$K \cite{andrewmarc}.  In an open and/or
high-baryon-density Universe, the level of anisotropy is
somewhat higher.  The signal in some of these models may be
detectable with planned interferometry
experiments \cite{andrewmarc}.

{\it Signatures of New Particle and Gravitational Physics:} The
CMB may be used as a probe of new particle physics in yet
another way:  One of the primary goals of experimental particle
physics these days is pursuit of a nonzero neutrino mass.  Some
recent (still controversial) experimental results suggest that
one of the neutrinos may have a mass of $\order(5\,{\rm eV})$
\cite{LSND}, and there have been some (again, still
controversial) arguments that such a neutrino mass is just what
is required to explain some apparent discrepancies between
large-scale-structure observations and the simplest
inflation-inspired standard-CDM model \cite{primack}.

If the neutrino does indeed have a mass of $\order(5 \, {\rm
eV})$, then roughly 30\% of the mass in the Universe will be in
the form of light neutrinos.  These neutrinos will affect the
growth of gravitational-potential wells near the epoch of last
scatter, and they will thus leave an imprint on the CMB angular
power spectrum \cite{dodelson}.  The effect of a light neutrino
on the power spectrum is small, so other cosmological parameters
that might affect the shape of the power spectrum at larger
$l$'s must be known well.  Still, Ref. \cite{eht} argues
that by combining measurements of the CMB power spectrum with
those of the mass power spectrum measured by, e.g., the Sloan
Digital Sky Survey, a neutrino mass of $\order(5 \, {\rm eV})$
can be determined. 

The CMB may also conceivably be used to test alternative gravity 
theories \cite{liddle,xueleijbd} such as Jordan-Brans-Dicke or more
general scalar-tensor theories.  The idea here is that in such a 
theory, the expansion rate at the epoch of last scatter will be
different, and this will provide a unique signature in the CMB
power spectrum.

{\it Cross-Correlation with the x-ray background:}  
In~an~Einstein-de~Sitter~Universe, large-angle CMB anisotropies
are produced by gravitational-potential differences induced by
density perturbations at the surface of last scatter at a redshift
$z\simeq1100$ via the Sachs-Wolfe (SW) effect.  In a flat
cosmological-constant Universe \cite{kofman}, or in an open
Universe \cite{kamspergelsug}, additional anisotropies are produced by
density perturbations at lower redshifts along the line of sight
via the integrated Sachs-Wolfe (ISW) effect.  Crittenden and Turok
\cite{ct} thus argued  that in a flat cosmological-constant
Universe, there should be some cross-correlation
between the CMB and a tracer of the mass distribution at low
redshifts; a similar cross-correlation should also arise in an
open Universe \cite{marc}.  

The x-ray background (XRB) currently offers perhaps the best tracer of 
the mass distribution out to redshifts of a few.  Boughn,
Crittenden, and Turok \cite{bct} determined an upper limit to
the amplitude of the cross-correlation between the Cosmic
Background Explorer (COBE) CMB map and the first High-Energy
Astrophysical Observatory (HEAO I) map of the 2--20 keV XRB, and
used it to constrain $\Omega_0\ga0.3$ in a flat
cosmological-constant Universe.  More recently, the analogous
calculation has been carried out for an open Universe
\cite{alimarc}.  It is found that in an open Universe with a
nearly scale-invariant spectrum of primordial adiabatic
perturbations, $\Omega_0 \ga 0.7$.  This result makes
open-CDM models with $\Omega_0 \simeq0.3-0.4$ unlikely.

\section{Dark Matter and New Particles}

\subsection{The Local Dark-Matter Density}

The extent of the luminous disk of our Galaxy, the Milky Way, is
roughly 10 kpc, and we live about 8.5 kpc from the center.
Due to our location {\it in} it, the rotation curve of the Milky
Way cannot be determined with the same precision as that of an
external spiral galaxy, such as that shown in
Fig. \ref{rotationfigure}.  However, it is qualitatively the
same.  The circular speed rises linearly from zero at the center
and asymptotes to roughly 220 km~sec$^{-1}$ somewhere near our
own Galactocentric radius and remains roughly flat all the way
out to $\sim25$~kpc.  Although direct measurements of the
rotation curve are increasingly difficult at larger radii, the
orbital motions of satellites of the Milky Way suggest that the
rotation curve remains constant all the way out to radii of 50
kpc and perhaps even farther.  According to Newton's laws, the
rotation speed should fall as $v_c \propto r^{-1/2}$ at radii
greater than the extent of the luminous disk.  However, it is
observed to remain flat to much larger distances.  It therefore
follows that the luminous disk and bulge must be immersed
in an extended dark halo (or that Newton's laws are violated).

Our knowledge of the halo comes almost solely from this rotation
curve.  Therefore, we do not know empirically if the halo is
round, elliptical, or perhaps flattened like the disk.  However,
there are good reasons to believe that the halo should be much
more diffuse than the disk.  The disk is believed to be flat
because luminous matter can radiate photons and therefore
gravitationally collapse to a pancake-like structure.  On the
other hand, dark matter (by definition) cannot radiate photons.
There are also now empirical arguments which involve, e.g., the
shape of the distribution of gas in the Milky Way, which suggest
that the dark halo should be much more diffuse than the disk \cite{rob}.

Assuming that the halo is therefore nearly round, it must have a 
density distribution like
\begin{equation}
     \rho(r) = \rho_0 {r_0^2 + a^2 \over r^2 + a^2},
\label{eq:rho}
\end{equation}
where $r$ is the radius, $r_0\simeq8.5$~kpc is our distance from
the center, $a$ is a to-be-determined core radius of the halo,
and $\rho_0$ is
the local halo density.  Such a halo would give rise to a
rotation curve,
\begin{equation}
     v_h^2(r) = 4\pi G \rho_0 (r_0^2 +a^2)
     \left[1-\left({a\over r}\right)\tan^{-1}\left({a\over
     r}\right)\right],
\end{equation}
where $G$ is Newton's constant.  If we know the rotation speed
contributed by the halo at two points, we can determine $\rho_0$
and $a$.  At large radii, the rotation curve of the Milky Way is
supported entirely by this dark halo, so $v_h(r\gg10\,{\rm kpc})
\simeq 220$~km~sec$^{-1}$.  However, the rotation curve locally
is due in part to the disk, $v_c^2(r_0) =
v_d^2(r_0)+v_h^2(r_0)$.  The disk contribution to the local
rotation speed is somewhat uncertain but probably falls in the
range $v_d(r_0) \simeq 118-155$~km~sec$^{-1}$, which means that the
halo contribution to the local rotation speed is $v_h(r_0)
\simeq 150-185$~km~sec$^{-1}$.  Given the local and asymptotic
rotation speeds, we infer that the local halo density is $\rho_0
\simeq 0.3-0.5$~GeV~cm$^{-3}$.  The particles which make up the
dark halo move locally in the same gravitational potential well
as the Sun.  Therefore, the virial theorem tells us that they
must move with velocities $v\sim v_c \sim220$~km~sec$^{-1}$.
Additional theoretical arguments suggest that the velocity
distribution of these particles is locally nearly isotropic and
nearly a Maxwell-Boltzmann distribution, (although the
assumption of a Maxwell-Boltzmann distribution will not affect
the detection rates discussed below \cite{ali}).
To sum, application of
Newton's laws to our Galaxy tells us that the luminous disk and
bulge must be immersed in a dark halo with a local density
$\rho_0 \simeq 0.4$~GeV~cm$^{-3}$ and that dark-matter particles
(whatever they are) move with velocities comparable to the local
circular speed.  More careful investigations along these lines
show that similar conclusions are reached even if we allow for
the possibility of a slightly flattened halo or a radial
distribution which differs from that in Eq. (\ref{eq:rho}) \cite{ali}.

\subsection{Axions}

Although supersymmetric particles seem to get more attention in
the literature lately, we should not forget that the axion also
provides a well-motivated and promising alternative dark-matter
candidate \cite{axion}.  The QCD Lagrangian may be written
\begin{equation}
     {\cal L}_{QCD} = {\cal L}_{\rm pert} + \theta {g^2 \over 32
     \pi^2} G \widetilde{G},
\end{equation}
where the first term is the perturbative Lagrangian responsible
for the numerous phenomenological successes of QCD.  However,
the second term (where $G$ is the gluon field-strength tensor
and $\widetilde{G}$ is its dual), which is a consequence of
nonperturbative effects, violates $CP$.  However, we know
experimentally that $CP$ is not violated in the strong
interactions, or if it is, the level of strong-$CP$ violation is
tiny.  From constraints to the neutron electric-dipole moment,
$d_n \la 10^{-25}$ e~cm, it can be inferred that $\theta
\la 10^{-10}$.  But why is $\theta$ so small?  This is the
strong-$CP$ problem.

The axion arises in the Peccei-Quinn solution to the strong-$CP$
problem \cite{PQ}, which close to twenty years after it was proposed still
seems to be the most promising solution.  The idea is to
introduce a global $U(1)_{PQ}$ symmetry broken at a scale
$f_{PQ}$, and $\theta$ becomes a dynamical field which is the
Nambu-Goldstone mode of this symmetry.  
At temperatures below the QCD phase transition,
nonperturbative quantum effects break explicitly the symmetry
and drive $\theta\rightarrow 0$.  The axion is the
pseudo-Nambu-Goldstone boson of this near-global symmetry.  Its
mass is $m_a \simeq\, {\rm eV}\,(10^7\, {\rm GeV}/ f_a)$, and its
coupling to ordinary matter is $\propto f_a^{-1}$.

{\it A priori}, the Peccei-Quinn solution works equally well for
any value of $f_a$ (although one would generically expect it to
be less than or of order the Planck scale).  However, a variety
of astrophysical observations and a few laboratory experiments
constrain the axion mass to be $m_a\sim10^{-4}$ eV, to within a
few orders of magnitude.  Smaller masses would lead to an
unacceptably large cosmological abundance.  Larger masses
are ruled out by a combination of constraints from supernova
1987A, globular clusters, laboratory experiments, and a search
for two-photon decays of relic axions \cite{ted}.

One conceivable theoretical difficulty with this axion mass
comes from generic quantum-gravity arguments \cite{gravity}.  For
$m_a\sim10^{-4}$ eV, the magnitude of the explicit symmetry
breaking is incredibly tiny compared with the PQ scale, so the
global symmetry, although broken, must be very close to exact.
There are physical arguments involving, for example, the
nonconservation of global charge in evaporation of a black hole
produced by collapse of an initial state with nonzero global
charge, which suggest that  global symmetries should be violated
to some extent in quantum gravity.  When one writes down a
reasonable {\it ansatz} for a term in a low-energy effective
Lagrangian which might arise from global-symmetry violation at
the Planck scale, the coupling of such a term is found to be
extraordinarily small (e.g., $\la 10^{-55}$).  Of course,
we have at this point no predictive theory of quantum gravity,
and several mechanisms for forbidding these global-symmetry
violating terms have been proposed \cite{solutions}.  Therefore,
these arguments by no means ``rule out'' the axion solution.
In fact, discovery of an axion would provide much needed clues
to the nature of Planck-scale physics.

Curiously enough, if the axion mass is in the relatively small viable
range, the relic density is $\Omega_a\sim1$ and may therefore
account for the halo dark matter.  Such axions would be produced
with zero momentum by a misalignment mechanism in the early
Universe and therefore act as cold dark matter.  During the process of
galaxy formation, these axions would fall into the Galactic
potential well and would therefore be present in our halo with a
velocity dispersion near 270 km~sec$^{-1}$.

Although the interaction of axions with ordinary matter is
extraordinarily weak, Sikivie proposed a very clever method of
detection of Galactic axions \cite{sikivie}.  Just as the axion couples to
gluons through the anomaly (i.e., the $G\widetilde{G}$ term),
there is a very weak coupling of an axion to photons through the
anomaly.  The axion can therefore decay to two
photons, but the lifetime is $\tau_{a\rightarrow \gamma\gamma}
\sim 10^{50}\, {\rm s}\, (m_a / 10^{-5}\, {\rm eV})^{-5}$ which
is huge compared to the lifetime of the Universe and therefore
unobservable.  However, the $a\gamma\gamma$ term in the
Lagrangian is ${\cal L}_{a\gamma\gamma} \propto a {\vec E} \cdot
{\vec B}$ where ${\vec E}$ and ${\vec B}$ are the electric and
magnetic field strengths.  Therefore, if one immerses a resonant
cavity in a strong magnetic field, Galactic axions which pass
through the detector may be converted to fundamental excitations
of the cavity, and these may be observable \cite{sikivie}.  Such
an experiment is currently underway \cite{axionexperiment}.
They have already begun to probe part of the cosmologically
interesting parameter space (no, they haven't found anything
yet) and expect to cover most of the interesting region
parameter space in the next three years.  A related experiment,
which looks for 
excitations of Rydberg atoms, may also find dark-matter
axions \cite{rydberg}. 
Although the sensitivity of this technique should be
excellent, it can only cover a limited axion-mass range.

It should be kept in mind that there are no accelerator tests
for axions in the acceptable mass range.  Therefore, these
dark-matter axion experiment are actually our {\it only}
way to test the Peccei-Quinn solution.

\subsection{Weakly-Interacting Massive Particles}

Suppose that in addition to the known particles of the
standard model, there exists a new, yet undiscovered, stable (or
long-lived) weakly-interacting massive
particle (WIMP), $\chi$.  At temperatures
greater than the mass of the particle, $T\gg m_\chi$, the
equilibrium number density of such particles is $n_\chi \propto
T^3$, but for lower temperatures, $T\ll m_\chi$, the equilibrium
abundance is exponentially suppressed, $n_\chi \propto
e^{-m_\chi/T}$.  If the expansion of the Universe were so slow
that  thermal equilibrium was always maintained, the number of
WIMPs today would be infinitesimal.  However, the Universe is
not static, so equilibrium thermodynamics is not the entire story.

%
\begin{figure}[htbp]
\centerline{\psfig{file=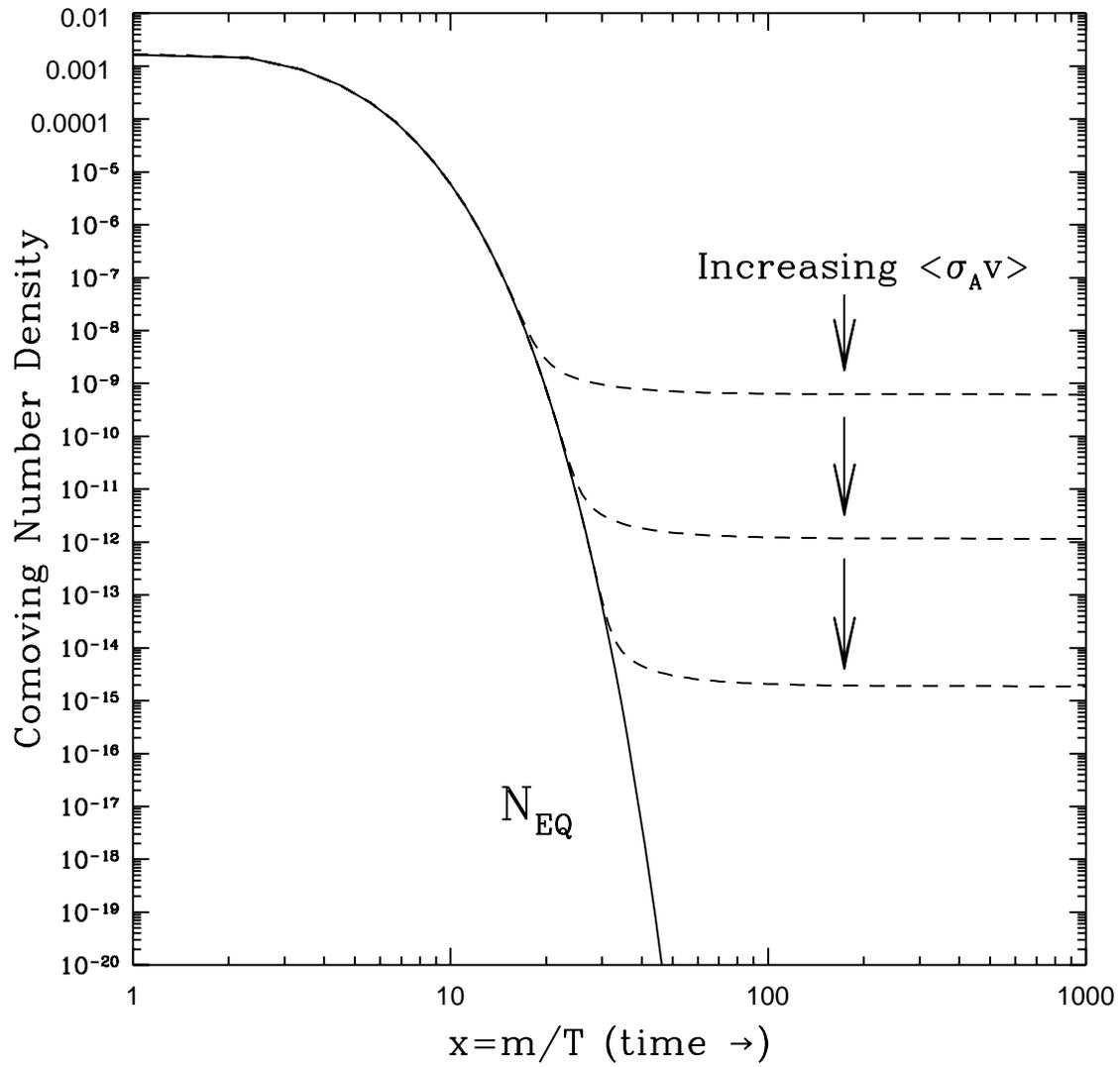,width=6in}}
\caption{Comoving number density of a WIMP in the early
	Universe.  The dashed curves are the actual abundance,
	and the solid curve is the equilibrium abundance.}
\label{YYY}
\end{figure}

At high temperatures ($T\gg m_\chi$), $\chi$'s are abundant and
rapidly converting to lighter particles and {\it vice versa}
($\chi\bar\chi\leftrightarrow l\bar l$, where $l\bar l$ are quark-antiquark and
lepton-antilepton pairs, and if $m_\chi$ is greater than the mass of the
gauge and/or Higgs bosons, $l\bar l$ could be gauge- and/or Higgs-boson
pairs as well).  Shortly after $T$ drops below $m_\chi$ the number
density of $\chi$'s drops exponentially, and the rate for annihilation of
$\chi$'s, $\Gamma=\VEV{\sigma v} n_\chi$---where $\VEV{\sigma v}$ is the
thermally averaged total cross section for annihilation of $\chi\bar\chi$
into lighter particles times relative velocity $v$---drops below 
the expansion rate, $\Gamma\la H$.  At this point, the $\chi$'s cease to
annihilate, they fall out of equilibrium, and a relic cosmological
abundance remains.

Fig.~\ref{YYY} shows numerical solutions to the Boltzmann
equation which determines the WIMP abundance.  The
equilibrium (solid line) and actual (dashed lines) abundances
per comoving volume are plotted as a
function of $x\equiv m_\chi/T$ (which increases with increasing time).
As the annihilation cross section is increased
the WIMPs stay in equilibrium longer, and we are left with a
smaller relic abundance.

An approximate solution to the Boltzmann equation yields the
following estimate for the current cosmological abundance of the
WIMP:
\begin{equation}
     \Omega_\chi h^2={m_\chi n_\chi \over \rho_c}\simeq
     \left({3\times 10^{-27}\,{\rm cm}^3 \, {\rm sec}^{-1} \over
     \sigma_A v} \right),
\label{eq:abundance}
\end{equation}
where $h$ is the Hubble constant in units of 100
km~sec$^{-1}$~Mpc$^{-1}$.  The result is to a first approximation
independent of the WIMP mass and is fixed primarily by its
annihilation cross section.

The WIMP velocities at freeze out are typically some appreciable
fraction of the speed of light.  Therefore, from
Eq.~(\ref{eq:abundance}), the WIMP will have a cosmological
abundance of order unity today if the annihilation cross section
is roughly $10^{-9}$ GeV$^{-2}$.  Curiously, this is the order
of magnitude one would expect from a typical electroweak cross
section, 
\begin{equation}
     \sigma_{\rm weak} \simeq {\alpha^2 \over m_{\rm weak}^2},
\end{equation}
where $\alpha \simeq {\cal O}(0.01)$ and $m_{\rm weak} \simeq
{\cal O}(100\, {\rm GeV})$.  The value of the cross section in
Eq.~(\ref{eq:abundance}) needed to provide $\Omega_\chi\sim1$
comes essentially from the age of the Universe.  However, there
is no {\it a priori} reason why this cross section should be of
the same order of magnitude as the cross section one would
expect for new particles with masses and interactions
characteristic of the electroweak scale.  In other words, why
should the age of the Universe have anything to do with
electroweak physics?  This ``coincidence'' suggests that if a
new, yet undiscovered, massive particle with electroweak
interactions exists, then it should have a relic density of
order unity and therefore provides a natural dark-matter
candidate.  This argument has been the driving force behind a
vast effort to detect WIMPs in the halo.

The first WIMPs considered were massive Dirac or Majorana
neutrinos with masses in the range of a few GeV to a few TeV.
(Due to the Yukawa coupling which gives a neutrino its mass, the
neutrino interactions become strong above a few TeV, and it no
longer remains a suitable WIMP candidate \cite{unitarity}.)  LEP ruled out
neutrino masses below half the $Z^0$ mass.  Furthermore, heavier
Dirac neutrinos have been ruled out as the primary component of
the Galactic halo by direct-detection experiments (described
below) \cite{heidelberg}, and heavier Majorana neutrinos have
been ruled out by indirect-detection
experiments \cite{kamiokande} (also described below) over much 
of their mass range.  Therefore, Dirac neutrinos cannot comprise
the halo dark matter \cite{griestsilk}; Majorana neutrinos can,
but only over a
small range of fairly large masses.  This was a major triumph
for experimental particle astrophysicists:\ the first
falsification of a dark-matter candidate.  However, theorists
were not too disappointed:  The stability of a fourth-generation
neutrino had to be postulated {\it ad hoc}---it was not
guaranteed by some new symmetry.  So although heavy neutrinos
were plausible, they certainly were not very well-motivated from
the perspective of particle theory.

{\it Supersymmetric Dark Matter:}
A much more promising WIMP candidate comes from supersymmetry
(SUSY) \cite{jkg,haberkane}.  SUSY was
hypothesized in particle physics to cure the naturalness problem
with fundamental Higgs bosons at the electroweak scale.
Coupling-constant unification at the GUT scale seems to be
improved with SUSY, and it seems to be an essential ingredient
in theories which unify gravity with the other three fundamental
forces.

As another consequence, the existence of a new symmetry,
$R$-parity, in SUSY theories guarantees that the lightest
supersymmetric particle (LSP) is stable.
In the minimal supersymmetric extension of the
standard model (MSSM), the LSP is usually the neutralino, a linear
combination of the supersymmetric partners of the photon, $Z^0$,
and Higgs bosons.  (Another possibility is the sneutrino, but
these particles interact like neutrinos and have been ruled out
over most of the available mass range \cite{sneutrino}.)  Given
a SUSY model, the cross section for
neutralino annihilation to lighter particles is straightforward,
so one can obtain the cosmological mass density.  The
mass scale of supersymmetry must be of order the weak scale to
cure the naturalness problem, and the neutralino will have only
electroweak interactions.  Therefore, it is to be expected that
the cosmological neutralino abundance is of order unity.  In
fact, with detailed calculations, one finds that the neutralino
abundance in a very broad class of supersymmetric extensions of
the standard model is near unity and can therefore account for
the dark matter in our halo \cite{ellishag}.  

\begin{figure}[htbp]
\centerline{\psfig{file=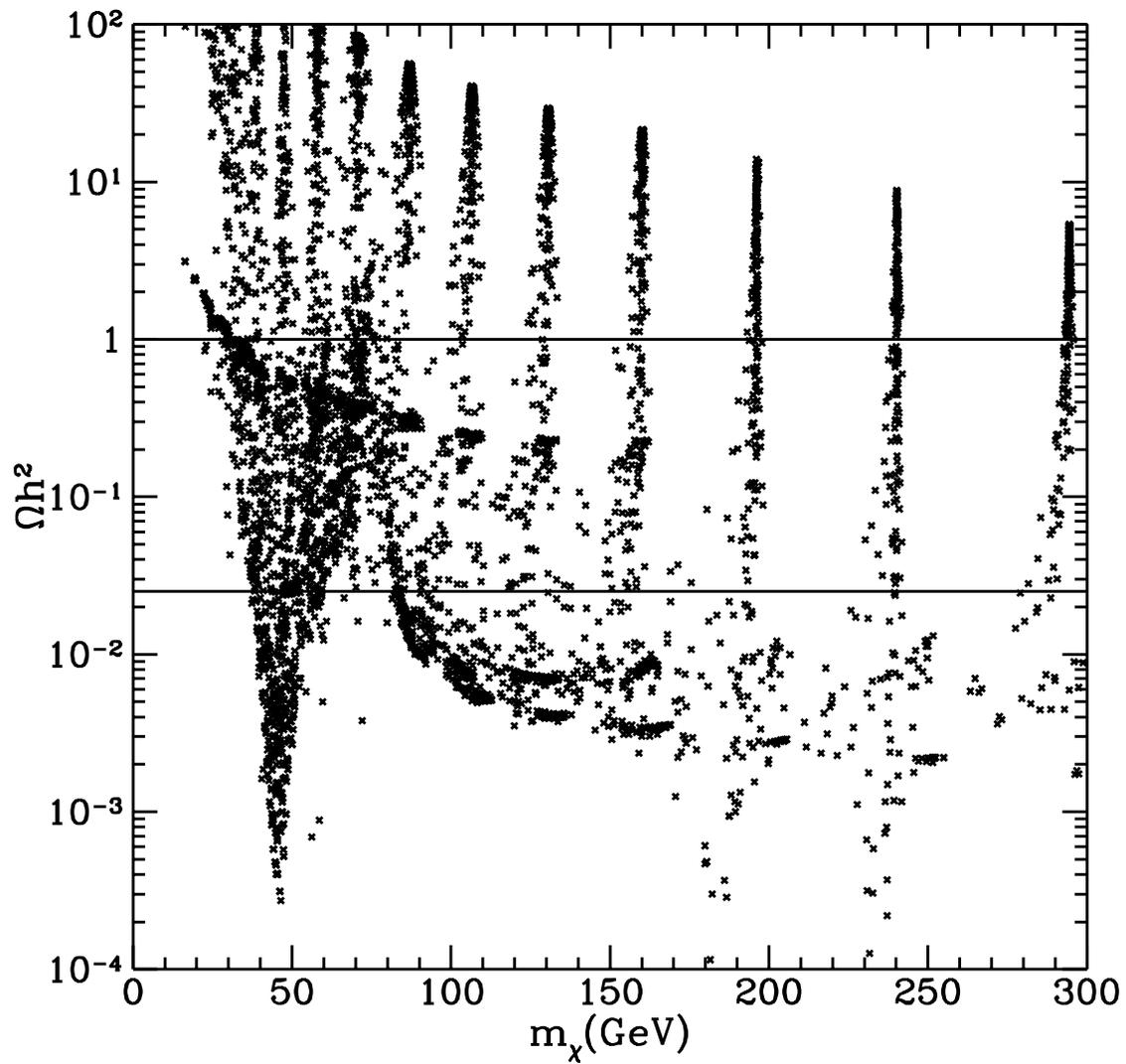,width=6in}}
\caption{Cosmological abundance of a WIMP versus the WIMP mass.
     Each point represents the result for a given choice of the
     MSSM parameters.  From Ref. \protect\cite{jkg}.}
\label{fig:relicabundance}
\end{figure}

This is
illustrated in Fig. \ref{fig:relicabundance} where the
cosmological abundance $\Omega_\chi$ (times $h^2$) is plotted
versus the neutralino mass $\Omega_\chi$.  Each point represent
one supersymmetric model, or equivalently, one choice of the
MSSM parameters.  Models with $\Omega_\chi h^2 \ga 1$ are
inconsistent with a conservative lower bound (10 Gyr) to the age 
of the Universe, and those with $\Omega_\chi h^2 \la 0.025$ are
cosmologically consistent, but probably too scarce to account
for the dark matter in galactic halos.  Still, numerous models
have an abundance between these two limits, and these models
make excellent dark-matter candidates.

{\it Direct Detection:} If neutralinos reside in the halo, there
are several avenues
toward detection \cite{jkg}.  One of the most promising
techniques currently being pursued involves searches for the
${\cal O}(10\, {\rm keV})$ recoils produced by elastic scattering of
neutralinos from nuclei in low-background
detectors \cite{witten,labdetectors}.  The idea here is simple.
A particle with mass $m_\chi\sim100$ GeV and electroweak-scale
interactions
will have a cross section for elastic scattering from a nucleus
which is $\sigma \sim 10^{-38}\,{\rm cm}^2$.  If the local halo
density is $\rho_0\simeq0.4$ GeV~cm$^{-3}$, and the particles
move with velocities $v\sim 300$ km~sec$^{-1}$, then the rate
for elastic scattering of these particles from, e.g., germanium
which has a mass $m_N \sim70$ GeV, will be $R \sim \rho_0
\sigma v / m_\chi/m_N \sim1$ event~kg$^{-1}$~yr$^{-1}$.  If a
$100$-GeV WIMP moving at $v/c\sim10^{-3}$ elastically scatters
with a nucleus of similar mass, it will impart a recoil energy
up to 100 keV to the nucleus.  Therefore, if we have 1 kg of
germanium, we expect to see roughly one nucleus per year
spontaneously recoil with an energy nearly 100 keV.

Of course, this is only a {\it very} rough calculation.  To do
the calculation more precisely, one needs to use a proper
neutralino-quark interaction, treat the QCD and
nuclear physics which takes you from a neutralino-quark
interaction to a neutralino-nucleus interaction, and integrate
over the WIMP velocity distribution.  Even if all of these
physical effects are included properly, there is still a
significant degree of uncertainty in the predicted event rates.
Although supersymmetry provides perhaps the most promising
dark-matter candidate (and solves numerous problems in particle
physics), it really provides little detailed predictive power.
In SUSY models, the standard-model particle spectrum is more
than doubled, and we really have no idea what the masses of all
these superpartners should be.  There are also couplings, mixing
angles, etc. Therefore, what theorists generally do is survey a
large set of models with masses and couplings within a
plausible range, and present results for relic abundances and
direct- and indirect-detection rates, usually as scatter plots
versus neutralino mass.

\begin{figure}[htbp]
\centerline{\psfig{file=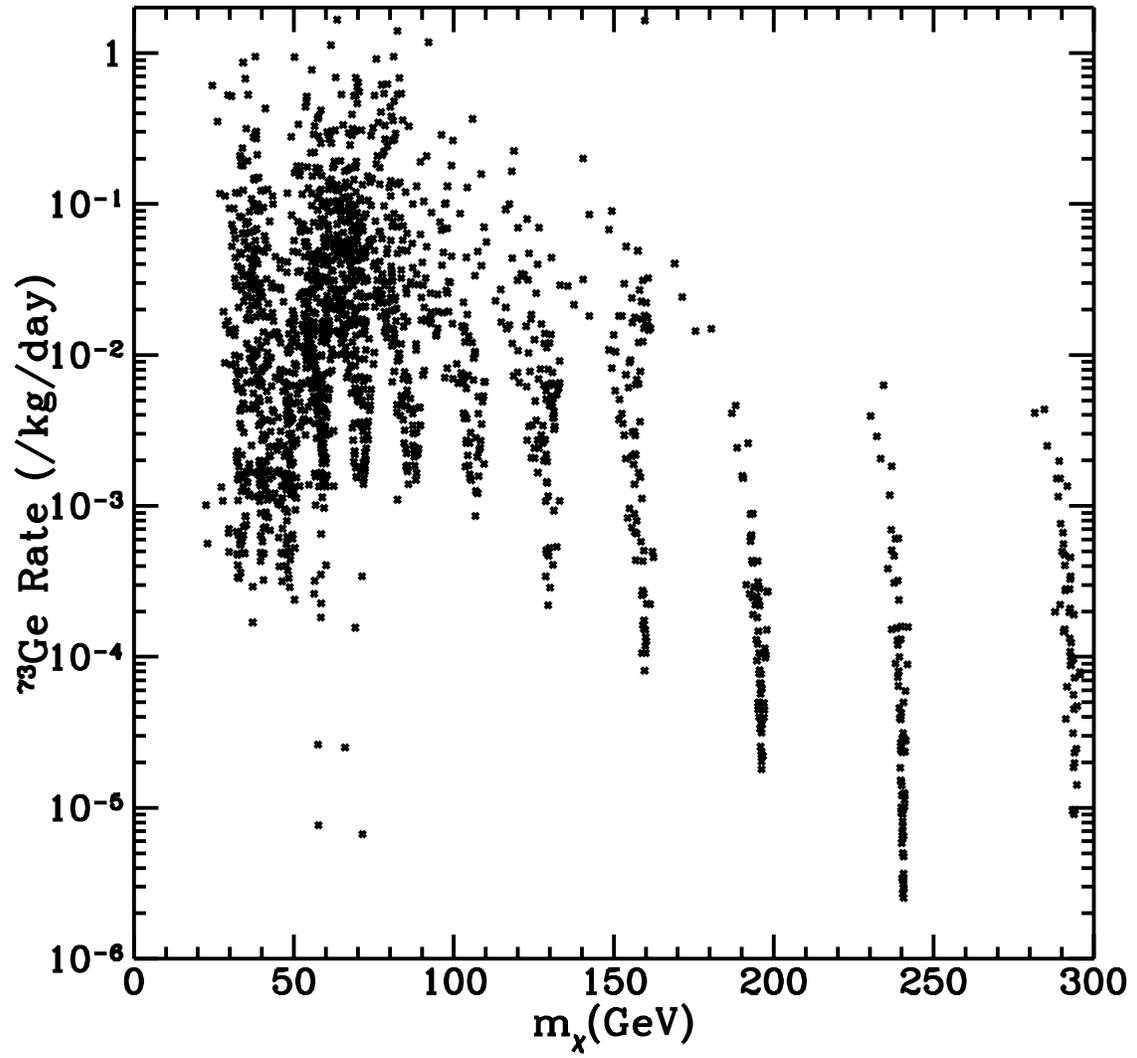,width=6in}}
\caption{Rates for direct detection of neutralinos of mass
     $m_\chi$ in a $^{73}$Ge detector.  From Ref. \protect\cite{jkg}.}
\label{fig:directdetection}
\end{figure}

After taking into account all the relevant physical effects and
surveying the plausible region of SUSY parameter space, one
generally finds that the predicted event rates seem to fall for
the most part between $10^{-4}$ to 10
events~kg$^{-1}$~day$^{-1}$ \cite{jkg}, as shown in
Fig. \ref{fig:directdetection} although again, there
may be models with higher or lower rates.  Current experimental
sensitivities in germanium detectors are around 10
events~kg$^{-1}$~day$^{-1}$ \cite{heidelberg}.  To illustrate
future prospects, consider the CDMS experiment \cite{cdms} which
expects to soon have a kg germanium detector with a background
rate of 1 event~day$^{-1}$.  After a one-year exposure, their
sensitivity would therefore be ${\cal O}(0.1\, {\rm
event~kg}^{-1}\,{\rm day}^{-1})$; this could be improved with
better background rejection.  Future detectors will achieve
better sensitivities, and it should be kept in mind that
numerous other target nuclei are being considered by other
groups.  However, it also seems clear that it will be quite a
while until a good fraction of the available SUSY parameter
space is probed.

{\it Indirect Detection:} Another strategy is observation of
energetic neutrinos produced
by annihilation of neutralinos in the Sun and/or Earth in converted
proton-decay and astrophysical-neutrino detectors (such as
MACRO, Kamiokande, IMB, AMANDA, and NESTOR) \cite{SOS}.  If, upon
passing through the Sun, a WIMP scatters elastically from a
nucleus therein to a velocity less than the escape velocity, it
will be gravitationally bound in the Sun.  This leads to a
significant enhancement in the density of WIMPs in the center of
the Sun---or by a similar mechanism, the Earth.  These WIMPs
will annihilate to, e.g., $c$, $b$, and/or $t$ quarks, and/or gauge and
Higgs bosons.  Among the decay products of these particles
will be energetic muon neutrinos which can escape from the
center of the Sun and/or Earth and be detected in neutrino
telescopes such as IMB, Kamiokande, MACRO, AMANDA, or NESTOR.
The energies of these muons will be typically 1/3 to 1/2 the
neutralino mass (e.g., 10s to 100s of GeV) so they will be much
more energetic---and therefore cannot be confused
with---ordinary solar neutrinos.  The signature of such a
neutrino would be the Cerenkov radiation emitted by an upward
muon produced by a charged-current interaction between the
neutrino and a nucleus in the rock below the detector.

The annihilation rate of these WIMPs is equal to the rate for
capture of these particles in the Sun.  This can be estimated in
order of magnitude by determining the rate at which halo WIMPs
elastically scatter from nuclei in the Sun.  The flux of
neutrinos at the Earth depends also on the Earth-Sun distance,
WIMP annihilation branching ratios, and the decay branching
ratios of the annihilation products.  The flux of upward muons
depends on the flux of neutrinos and the cross section for
production of muons, which depends on the square of the neutrino
energy.  

\begin{figure}[htbp]
\centerline{\psfig{file=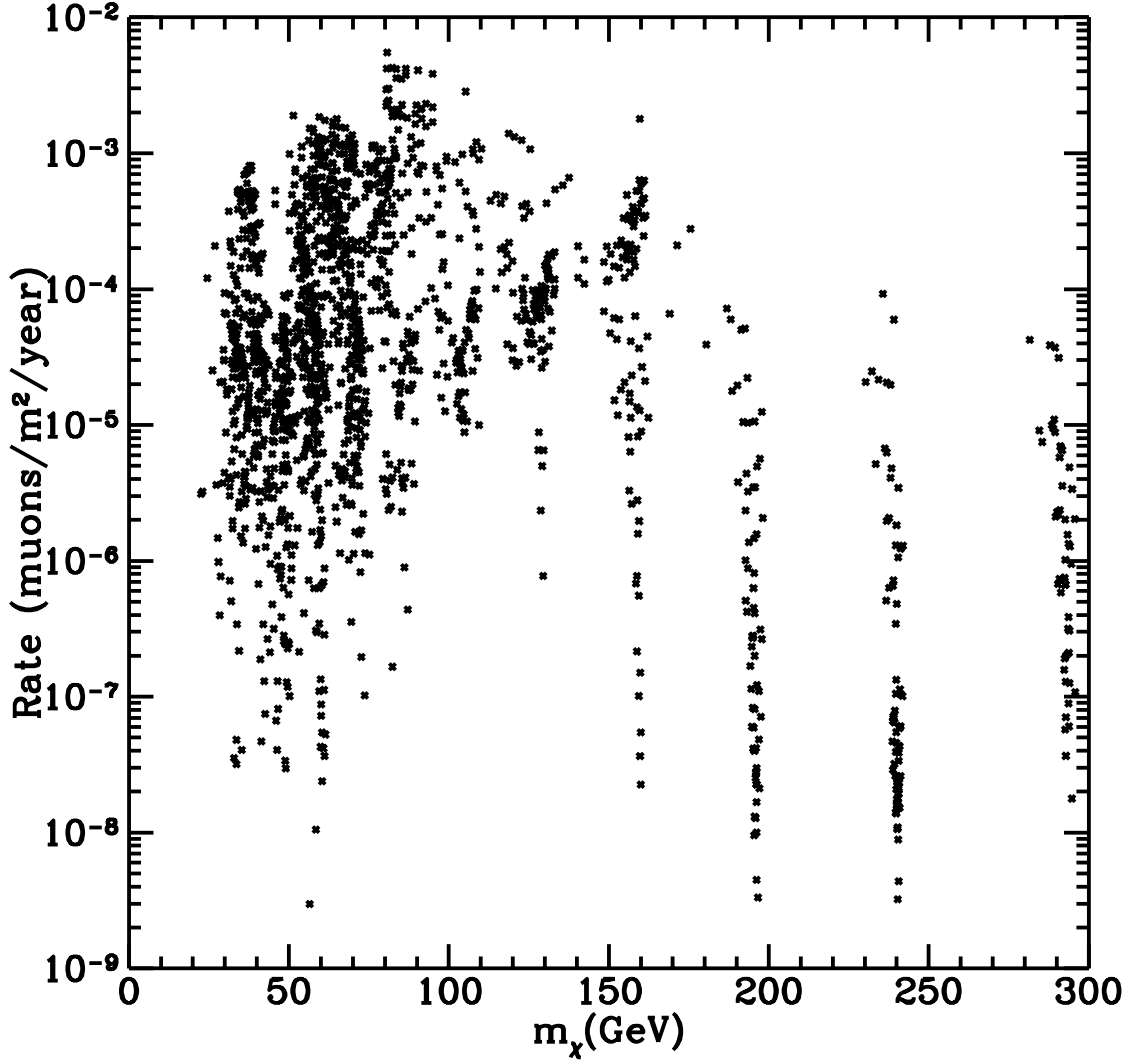,width=6in}}
\caption{Rates for indirect detection of neutralinos of mass
     $m_\chi$ via observation of energetic neutrinos from
     annihilation in the Sun and Earth.  From Ref. \protect\cite{jkg}.}
\label{fig:indirect}
\end{figure}

As in the case of direct detection, the precise
prediction involves numerous factors from particle and nuclear
physics and astrophysics, and on the SUSY parameters.
When all these factors are taken into account, predictions for
the fluxes of such muons in SUSY models
seem to fall for the most part between $10^{-6}$ and 1
event~m$^{-2}$~yr$^{-1}$ \cite{jkg}, as shown in
Fig. \ref{fig:indirect}, although the numbers may be a bit
higher or lower in some models.  Presently, IMB and Kamiokande
constrain the flux of energetic neutrinos from the Sun to be
less than about 0.02 m$^{-2}$~yr$^{-1}$ \cite{kamiokande,imb},
and the Baksan limit is perhaps a factor-of-two
better~\cite{baksan}.  MACRO expects to be
able to improve on this sensitivity by perhaps an order of
magnitude.  Future detectors may be able to improve even
further.  For example, AMANDA expects to have an area of
roughly $10^4$ m$^2$, and a $10^6$-m$^2$
detector is being discussed.  However, it should be kept in
mind that without muon energy resolution, the sensitivity of
these detectors will not approach the inverse exposure; it will
be limited by the atmospheric-neutrino background.  If a
detector has good angular resolution, the signal-to-noise ratio
can be improved, and even moreso with energy resolution, so
sensitivities approaching the inverse exposure could be
achieved~\cite{joakim}.  Furthermore, ideas for neutrino detectors with
energy resolution are being discussed~\cite{wonyong}, although
at this point these appear likely to be in the somewhat-distant future.

{\it Direct/Indirect Comparison:} With two promising avenues
toward detection, it is natural to
inquire which is most promising.  Due to the abundance of
undetermined SUSY parameters and the complicated dependence of
event rates on these parameters, the answer to this question is
not entirely straightforward.
Generally, most theorists have just plugged SUSY parameters
into the machinery which produces detection rates and plotted
results for direct and indirect detection.  However, another
approach is to compare, in a somewhat model-independent although
approximate fashion, the rates for direct and indirect
detection~\cite{jkg,taorich,bernard}.  The underlying
observation is that the rates for the 
two types of detection are both controlled primarily by the WIMP-nucleon
coupling.  One must then note that WIMPs generally undergo one
of two types of interaction with the nucleon: an axial-vector
interaction in which the WIMP couples to the nuclear spin
(which, for nuclei with nonzero angular momentum is roughly 1/2
and {\it not} the total angular momentum), and a scalar
interaction in which the WIMP couples to the total mass of the
nucleus.  The direct-detection rate depends on the WIMP-nucleon
interaction strength and on the WIMP mass.  On the other hand,
indirect-detection rates will have an additional dependence on
the energy spectrum of neutrinos from WIMP annihilation.  By
surveying the various possible neutrino energy spectra, one
finds that for a given neutralino mass and annihilation rate in
the Sun, the largest upward-muon flux is roughly three times as
large as the smallest~\cite{bernard}.  So even if we assume the
neutralino-nucleus interaction is purely scalar or purely
axial-vector, there will still be a residual model-dependence of
a factor of three when comparing direct- and indirect-detection
rates.

For example, for scalar-coupled WIMPs, the event rate in a kg
germanium detector will be
equivalent to the event rate in a $(2-6)\times 10^6$ m$^2$
neutrino detector for 10-GeV WIMPs and $(3-5)\times10^4$ m$^2$
for TeV WIMPs~\cite{bernard}.  Therefore, the relative
sensitivity of indirect detection when compared with the
direct-detection sensitivity increases with mass.
The bottom line of such an analysis seems to be that
direct-detection experiments will be more sensitive to
neutralinos with scalar interactions with nuclei, although
very-large neutrino telescopes may achieve comparable
sensitivities at larger WIMP masses.  This should
come as no surprise given the fact that direct-detection
experiments rule out Dirac neutrinos \cite{heidelberg}, which
have scalar-like interactions, far more effectively than
do indirect-detection experiments \cite{bernard}.

Generically, the sensitivity of
indirect searches (relative to direct searches) should be better
for WIMPs with axial-vector interactions, since the Sun is
composed primarily of nuclei with spin (i.e., protons).
However, a comparison of direct-
and indirect-detection rates is a bit more difficult for
axially-coupled WIMPs, since the nuclear-physics uncertainties
in the neutralino-nuclear cross section are much greater, and
the spin distribution of each target nucleus must be modeled.
Still, in a careful analysis, Rich and Tao found that in 1994,
the existing sensitivity of energetic-neutrino searches to
axially-coupled WIMPs greatly exceeded the sensitivities of
direct-detection experiments \cite{taorich}.

To see how the situation may change with future
detectors, let us consider a specific axially-coupled
dark-matter candidate, the light Higgsino recently put forward by
Kane and Wells \cite{kanewells}.  Even if this candidate is
inconsistent, it will serve as a toy model for a WIMP with
primarily spin-dependent interactions.  In order to explain the
anomalous CDF $ee\gamma \gamma + \slashchar{E}_T$ \cite{CDF},
the $Z\rightarrow b\bar b$ anomaly,
and the dark matter, this Higgsino must have a mass between
30--40 GeV.  Furthermore, the coupling of this Higgsino to
quarks and leptons is due primarily to $Z^0$ exchange with a
coupling proportional to $\cos 2\beta$, where $\tan\beta$ is the
usual ratio of Higgs vacuum expectation values in
supersymmetric models.  Therefore, the usually messy cross
sections one deals with in a general MSSM simplify for this
candidate, and the cross sections needed for the cosmology of
this Higgsino depend only on the two parameters $m_\chi$ and
$\cos2\beta$.  Furthermore, since the neutralino-quark
interaction is due only to $Z^0$ exchange, this Higgsino will
have only axial-vector interactions with nuclei.

The Earth is composed primarily of spinless nuclei, so WIMPs
with axial-vector interactions will not be captured in the Earth,
and we expect no neutrinos from WIMP annihilation therein.
However, most of the mass in the Sun is composed of nuclei with
spin (i.e., protons).  The flux of upward muons induced by
neutrinos from
annihilation of these light Higgsinos would be $\Gamma_{\rm
det}\simeq 2.7\times10^{-2}\, {\rm m}^{-2}\, {\rm yr}^{-1}\,
\cos^2 2\beta$ \cite{katie}.  On the other hand, the rate for scattering from
$^{73}$Ge is $R\simeq 300\, \cos^2 2\beta\, {\rm kg}^{-1}\,
{\rm yr}^{-1}$ \cite{kanewells,katie}.  For illustration, in
addition to their kg of
natural germanium, the CDMS experiment also plans to
run with 0.5 kg of (almost) purified $^{73}$Ge.  With a
background event rate of roughly one event~kg$^{-1}$~day$^{-1}$,
after one year, the $3\sigma$ sensitivity of the experiment will
be roughly 80 kg$^{-1}$~yr$^{-1}$.  Comparing the predictions
for direct and indirect detection of this axially-coupled WIMP,
we see that the enriched-$^{73}$Ge sensitivity should improve on
the {\it current} 
limit to the upward-muon flux ($0.02$ m$^{-2}$ yr$^{-1}$)
roughly by a factor of 4.  When we compare this with the forecasted
factor-of-ten improvement expected in MACRO, it appears that the
sensitivity of indirect-detection experiments looks more
promising.  Before
drawing any conclusions, however, it should be noted that the
sensitivity in detectors with other nuclei with spin may be
significantly better.  On the other hand, the sensitivity of
neutrino searches increases relative to direct-detection
experiments for larger WIMP masses.  It therefore seems at this
point that the two schemes will be competitive for detection of
light axially-coupled WIMPs, but the neutrino telescopes may
have an advantage in probing larger masses.

A common question is whether
theoretical considerations favor a WIMP which has predominantly
scalar interactions or whether they favor axial-vector
couplings.  Unfortunately, there is no
simple answer.  When detection of supersymmetric dark matter was
initially considered, it seemed that the neutralino in most
models would have predominantly axial-vector interactions.  It
was then noted that in some fraction of models where the
neutralino was a mixture of Higgsino and gaugino, there could be
some significant scalar coupling as well \cite{kim}.  As it became evident
that the top quark had to be quite heavy, it was realized that
nondegenerate squark masses would give rise to scalar couplings
in most models \cite{drees}.  However, there are still large regions of
supersymmetric parameter space where the neutralino has
primarily axial-vector interactions, and in fact, the Kane-Wells
Higgsino candidate has primarily axial-vector interactions.  The
bottom line is that theory cannot currently reliably say which
type of interaction the WIMP is likely to have, so
experiments should continue to try to target both.

\begin{figure}[htbp]
\centerline{\psfig{file=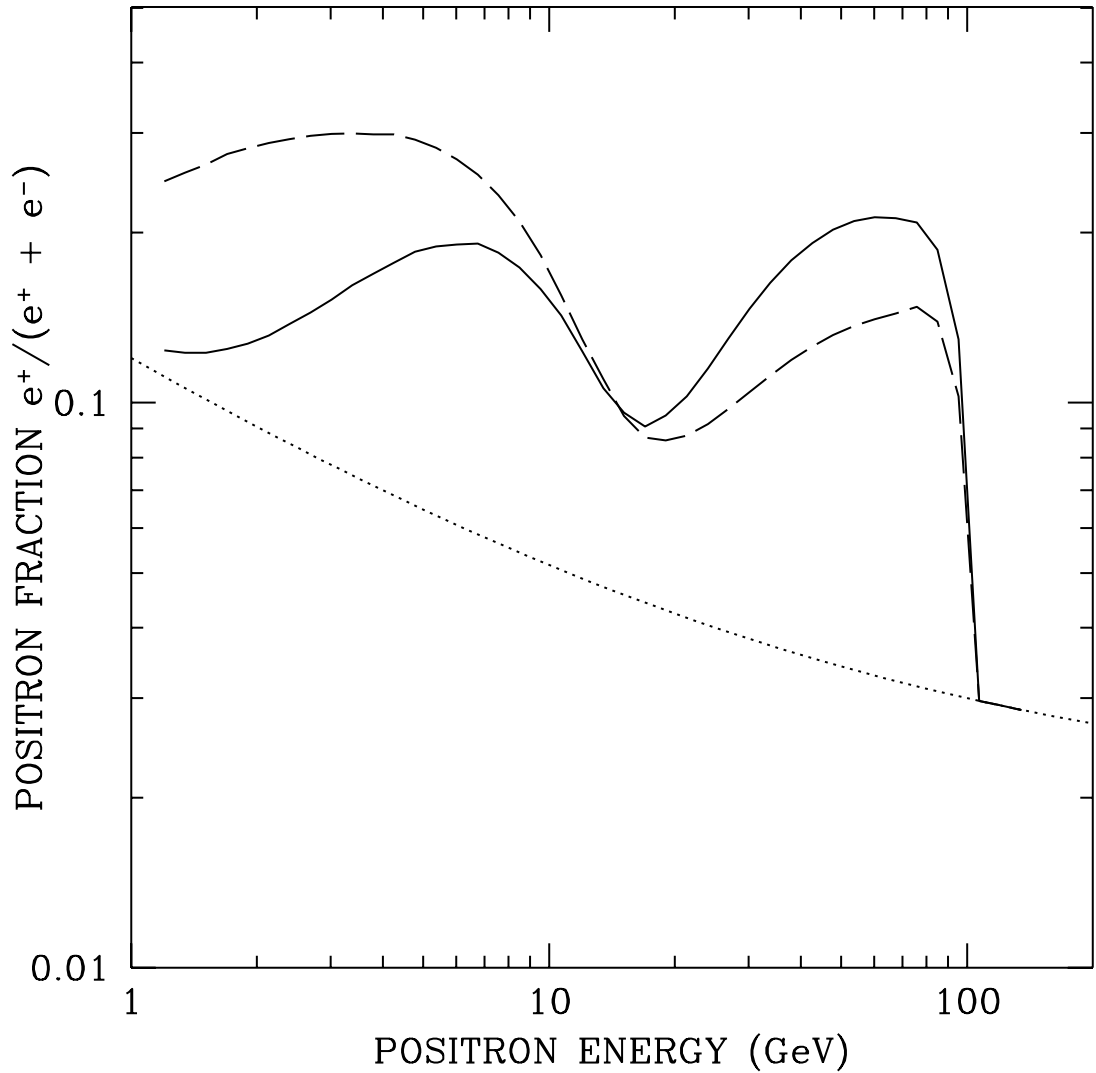,width=6in}}
\caption{The differential positron flux divided by the sum of
     the differential electron-plus-positron flux as a function
     of energy for a neutralino of mass 120 GeV, for two
     different models of cosmic-ray propagation.  The dotted
     curve is the background expected from traditional sources.
     From Ref. \protect\cite{positrons}.}
\label{fig:positrons}
\end{figure}

\begin{figure}[htbp]
\centerline{\psfig{file=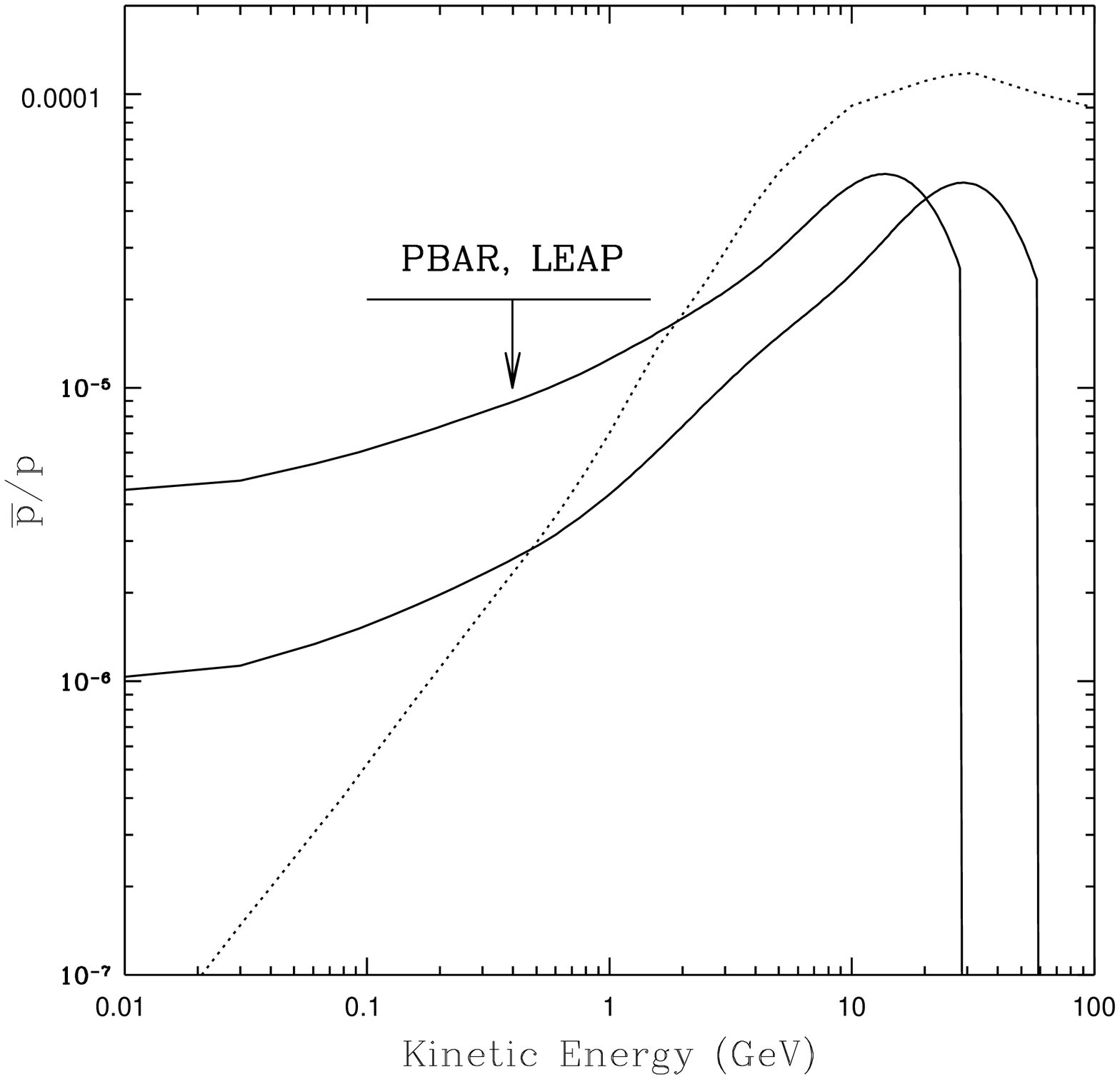,width=6in}}
\caption{The differential cosmic-ray antiproton flux divided by
     the proton flux as a function of energy for a neutralino of 
     mass 30 GeV and a neutralino of mass 60 GeV.  The dotted
     curve is the background expected from traditional sources.
     Ref. \protect\cite{antiprotons}.}
\label{fig:pbar}
\end{figure}

{\it Cosmic Rays from WIMP Annihilation in the Galactic Halo:}
WIMPs may also be detected via observation of anomalous
cosmic-ray positrons, antiprotons, and gamma rays produced by
WIMP annihilation in the Galactic halo.  The difficulty
in inferring the existence of particle dark matter from cosmic
rays lies in discrimination between WIMP-induced cosmic rays and
those from standard ``background'' sources.  However, WIMPs may
produce distinctive cosmic-ray signatures.  As illustrated in
Fig. \ref{fig:positrons}, WIMP annihilation will produce a
cosmic-ray-positron excess at high energies that cannot be
mimicked by any traditional astrophysical source.  Similarly,
WIMP annihilation will produce an antiproton excess at low
energies that is difficult to explain with traditional
astrophysical sources, as illustrated in Fig. \ref{fig:pbar}.
Direct annihilation of two
WIMPs to two photons will produce a gamma-ray line at an energy
equal to the WIMP mass.  No other imaginable astrophysical
mechanism could produce a monoenergetic gamma-ray signal at such 
an energy.  It should be kept in mind, however, that due to
several astrophysical uncertainties, it is difficult to make
reliable predictions for a given particle dark-matter candidate, 
so negative results from cosmic-ray searches cannot generally be
used to constrain dark-matter candidates.  On the other hand,
if observed, these cosmic-ray signatures could provide a
smoking-gun signal for the existence of WIMPs in the halo.

\section{Discussion}

We are likely on the verge of great discoveries in cosmology.
MAP and Planck will provide cosmological data of unprecedented
precision that should clarify the origin of large-scale
structure, and the values of several crucial cosmological
parameters.  

If MAP and Planck find a CMB temperature-anisotropy spectrum
consistent with a flat Universe and nearly--scale-free
primordial adiabatic perturbations, then the next step will be
to isolate the gravity waves with the polarization of the CMB.
If inflation has something to do with grand unification, then it
is possible that Planck's polarization sensitivity will be
sufficient to see the polarization signature of gravity waves.
However, it is also quite plausible that the height of the
inflaton potential may be low enough to elude detection by
Planck.  If so, then a subsequent experiment with better
sensitivity to polarization will need to be done.

Since big-bang nucleosynthesis
predicts that the baryon density is $\Omega_b \la 0.1$ and
inflation predicts $\Omega=1$, another prediction of inflation
is a significant component of nonbaryonic dark matter.  This can
be either in the form of vacuum energy (i.e., a cosmological
constant), and/or some new elementary particle.  Therefore,
discovery of particle dark matter could be interpreted as
evidence for inflation.

Axions and WIMPs have not only intrigued theorists; a large
community of experimentalists have devoted themselves to finding
these particles.  However, it should also be emphasized that
although very attractive, these are still speculative ideas.
There is still no direct evidence in accelerator experiments or
otherwise for the existence of axions or of supersymmetry.  The
dark matter could be composed of something completely
different.  However, as argued here, the evidence for the
existence of nonbaryonic dark matter is indeed extremely
compelling, and the two particles discussed here provide our
most promising candidates.  Although it provides an enormous
experimental challenge, it is clear that discovery of particle
dark matter would be truly revolutionary for both particle
physics and cosmology.

Although most of the calculations needed for predicting rates
for detection of supersymmetric dark matter have been completed
and are amalgamated in Ref. \cite{jkg}, there continue to be new
more general and/or more precise calculations.  For example,
just in the past year, there has been the first complete
calculation of the cross section for neutralino annihilation to
two photons  \cite{ullioone} as well as the first calculation of
the cross section for neutralino annihilation to a photon and a
$Z^0$ boson, and these will be needed for accurate calculations
of the strength of the photon line from WIMP annihilation in the 
Galactic halo.  The possible effects of CP violation in the
neutralino mass matrix have begun to be explored \cite{falk}.
Some non-negligible three-body final states from neutralino
annihilation have recently been discussed, and these may be
important for accurate calculations of indirect-detection rates
in certain regions of parameter space \cite{3body}.

Before closing, it should be noted that there is even more that
can be learned from the CMB and more ways to test and probe the
physics of inflation.
For example, inflation also predicts that the distribution of primordial
density perturbations is gaussian, and this can be tested with
CMB temperature maps and with the study of the large-scale
distribution of galaxies.  

As another example, large-scale galaxy surveys will soon map the
distribution of mass in the Universe today, and CMB experiments
will shortly determine the mass distribution in the early
Universe.  The next step will be to fill in the precise history
of structure formation in the ``dark ages'' after recombination
but before redshifts of a few.  Reconstruction of this epoch of
cosmic history will likely require amalgamation of the
complementary information provided by a number observations in
several wavebands.  Detection of the secondary CMB anisotropies
at arcminute scales produced by scattering from reionized clouds
will provide an indication of the epoch of reionization, and
therefore the epoch at which structures first undergo
gravitational collapse in the Universe.

\section*{Acknowledgments}
This work was supported by the U.S. D.O.E. Outstanding Junior
Investigator Award under contract DEFG02-92-ER 40699, NASA
NAG5-3091, and the Alfred P. Sloan Foundation.

\end{document}